\documentclass[preprint2]{aastex6}

\shorttitle{SNe I\lowercase{b} and II\lowercase{b} progenitors in binary systems}
\shortauthors{Yoon, Dessart \& Clocchiatti}

\begin{document}


\title{Type I\lowercase{b} and II\lowercase{b} supernova progenitors in interacting binary systems}

\author{Sung-Chul Yoon\altaffilmark{1}}  
\affil{ Department of Physics and Astronomy, Seoul National University, Gwanak-ro 1, Gwanak-gu, Seoul, 08826, Korea}
\altaffiltext{1}{yoon@astro.snu.ac.kr}
\author{Luc Dessart}
\affil{Unidad Mixta Internacional Franco-Chilena de Astronom{\'i}a (CNRS UMI 3386), Departamento de Astronom{\'i}a, Universidad de Chile, Camino El Observatorio 1515, Las Condes, Santiago, Chile}
\and
\author{Alejandro Clocchiatti}
\affil{Instituto de Astrof\'{\i}sica, Pontificia Universidad Cat\'olica de Chile and Millennium Institute of Astrophysics, Chile}

\begin{abstract}
We explore properties
of Type Ib and IIb SN progenitors that are produced by stable mass
transfer in binary systems, using a new grid of stellar evolution models from
an initial primary mass in the range of 10 - 18~$\mathrm{M_\odot}$ at solar and
Large Magellanic Cloud metallicities. We find that blue and yellow supergiant
SN IIb progenitors (e.g., of SN 2008ax, SN 2011dh, SN 2016gkg) have
a hydrogen envelope mass less than about 0.15~$\mathrm{M_\odot}$, 
mostly resulting from early Case B mass transfer with relatively 
low initial masses and/or low metallicity. 
Red supergiant  (RSG) SN IIb progenitors (e.g., of SN 1993J, SN 2013df) are
produced via late Case B mass transfer and  have a
more massive hydrogen envelope ($M_\mathrm{H,env} > 0.15~\mathrm{M_\odot}$). 
SN Ib progenitors are predominantly produced by early Case B mass
transfer.  Our models predict that SN IIb progenitors are systematically more
luminous in the optical ($-8.0 \lesssim M_\mathrm{V} \lesssim -5.0$) than the
majority of SN Ib progenitors ($M_\mathrm{V} \gtrsim -5.0$) for our considered
initial mass range. However, the optically bright progenitor of SN Ib iPTF13bvn
(i.e., $M_\mathrm{V} \simeq -6.5$) can be well explained by a
relatively low-mass progenitor with a final mass of $\sim
3.0~\mathrm{M_\odot}$.  The event rate of blue and yellow SN IIb progenitors
would increase as metallicity decreases, while the event
rate of SN Ib progenitors would decrease instead. By contrast, the population
of RSG SN IIb progenitors would not be significantly affected by
metallicity.  
\end{abstract}


\section{Introduction}

Type IIb supernovae (SNe) belong to the class of hydrogen-poor
core-collapse SNe.  They appear as a Type II SN initially but gradually turn
into a Type Ib SN days to weeks after the explosion.  This property can be best
explained by a progenitor star having a small amount of hydrogen
($M_\mathrm{H,env} \approx 0.01 - 1.0~ \mathrm{M_\odot}$) in the envelope.  

The inferred pre-explosion masses of Type IIb SN (hereafter, SN IIb) progenitors are typically about 3
- 6~$\mathrm{M_\odot}$ (e.g., SN 1993J, \citealt{Woosley94, Shigeyama94},
  \citealt{Blinnikov98}; SN 2008ax, \citealt{Taubenberger11, Folatelli15}; SN 2011dh,
\citealt{Bersten12, Ergon15}; SN 2011fu, \citealt{Kumar13};  SN 2011ei,
\citealt{Milisavljevic13}; SN 2011hs, \citealt{Bufano14}; SN 2013df, \citealt{Szalai16}).
This implies that their
immediate progenitors are not massive Wolf-Rayet (WR) stars of WNh type, which
are more massive than 8 - 10~$M_\odot$, and that the initial mass of a SN
IIb progenitor is in the range of 10 - 18~$\mathrm{M_\odot}$.  It is therefore
most likely that SN IIb progenitors have their hydrogen envelope
mainly removed by Roche-lobe overflow rather than by stellar wind mass
loss~\citep{Ensman88, Podsiadlowski93, Stancliffe09,Claeys11,Dessart11,  
Bersten12,  Benvenuto13, Dessart15, Folatelli15, Dessart16}.

The progenitors of a few SNe IIb have been directly identified in pre-SN images
as supergiants with a radius $\gtrsim$ 200~$\mathrm{R_\odot}$  (i.e., SN 1993J,
SN 2011dh, and SN 2013df; \citealt{Aldering94, Maund04, Maund11, VanDyk11,
VanDyk14}).  Comparison with stellar evolution and supernova models implies
that these progenitors had an inflated hydrogen envelope of a small mass.   The
luminosities of these progenitors are in the range of $\log L/\mathrm{L_\odot}
= 4.92 - 5.12$. The corresponding intial mass is in the range of 11 -
17~$\mathrm{M_\odot}$, which is consistent with the inferred pre-explosion
masses of 3 - 6~$\mathrm{M_\odot}$.  The inflated hydrogen envelope of supergiant
  progenitors significantly influences the early-time light curves of SNe IIb
during the shock-cooling phase. In particular, several studies conclude that
the sustained high luminosity at early times observed in some SNe IIb is
related to this inflated hydrogen envelope \citep[e.g., SN 1993J, SN 2011hs, SN
2011fu \& SN 2013df;][]{Bersten12, Nakar14, Morozova15}. 

However, observations indicate that not all SNe IIb have a supergiant
progenitor. For example, very weak or no sign of sustained brightness at early
times before the rise to the $^{56}$Ni-powered peak in the light curve is found
for SN  1996cb and 2008ax~\citep{Qiu99,Pastorello08}, which implies their
progenitors were fairly compact.  The analysis of SN 2008ax by
\citet{Folatelli15} indicates that its progenitor had a radius of $R \lesssim
50~\mathrm{R_\odot}$ and a pre-explosion mass of about 4 --
5~$\mathrm{M_\odot}$.  Evolutionary models with initial masses of 13 -
18~$\mathrm{M}_\odot$ also indicate that solutions for compact SN IIb
progenitors ($R \lesssim 50 ~\mathrm{M_\odot}$) can be obtained with binary
systems of relatively short initial periods ($P_\mathrm{init} \lesssim \sim
10~\mathrm{d}$; \citealt{Yoon10, Folatelli15}), while red-supergiant
progenitors can be produced with much longer initial periods ($P_\mathrm{init}
= \sim 1000~\mathrm{d}$; \citealt{Podsiadlowski93, Stancliffe09,Claeys11}).  A
yellow supergiant progenitor could be obtained with an initial period of $\sim
100$~d~\citep{Stancliffe09, Bersten12, Benvenuto13}.  The
observational diversity of the signatures of interaction with circumstellar
matter for some SNe IIb may also reflect different histories of mass loss due
to different progenitor properties \citep[e.g.,][]{Chevalier10, BenAmi15,
Maeda15,Kramble16}. 

The stellar evolution studies quoted above mostly focused on explaining
specific cases for SN IIb progenitors.  The purpose of this study is to have a
more comprehensive view on the properties of SN IIb progenitors and to discuss
the observational diversity of SNe IIb.  For this purpose, we present a new
grid of binary star evolution covering a range of initial primary-star mass (10
-- 18~$\mathrm{M_\odot}$) and a range of initial period ($\sim$~10 -- 3000~d)
for two different metallicities (solar and Large Magellanic Cloud values).
Although our discussion is focused on SN IIb progenitors, SNe IIb and Ib form a
continuous sequence of the increasing loss of the hydrogen-rich envelope and
therefore represent a unique family (distinct from SNe Ic).  In this context,
our model grid includes SNe Ib progenitor models for which no hydrogen is left
in the envelope.   We compare their properties at the final stage with those of
SN IIb progenitors. This complements our previous work on the properties of SN
Ib/c progenitors~\citep{Yoon10, Yoon12, Kim15}.

This paper is organized as follows.  We present the numerical method and
physical assumptions adopted for our evolutionary models in
Section~\ref{sect:method}. The main results are presented in
Section~\ref{sect:results}, The properties of SN IIb and SN Ib progenitors at
the final stage are discussed in Sections~\ref{sect:snIIb} and~\ref{sect:snIb},
respectively.  The effect of metallicity on progenitor properties is addressed
in Section~\ref{sect:metallicity}.  We conclude our study in
Section~\ref{sect:conclusions}. 

\section{Numerical method and physical assumptions}\label{sect:method}

The model grid presented in this study has been calculated with the MESA
code~~\citep{Paxton11, Paxton13, Paxton15}. We adopt  the Schwarzschild
criterion for convection. We consider convective overshoot using a step
function over a layer of thickness $l_\mathrm{OV} = 0.3~H_P$ above the hydrogen
burning convective core,  where $H_P$ is the pressure scale height at the outer
boundary of the core.  We use the Dutch scheme in MESA for both hot and cool
wind mass loss rates, with the Dutch scaling factor of 1.0. 

The energy transport efficiency by convection is usually parameterized by the
mixing length $l_\mathrm{m}$ in units of $H_P$.  We
adopt the default value of mesa ($l_\mathrm{m} = 1.5 H_P$).  We do not employ
the artificial boost of convective energy transport using an excessive
temperature gradient in the convective envelope that was introduced by
~\citet{Paxton13}. 

We employ the optically thick mass transfer prescription by \citet{Kolb90}
for calculating the mass transfer rate. This is well suited for the case of
mass transfer from a supergiant star whose pressure scale height in the outer
layers may become comparable to the size of the star~\citep{Pastetter89}. 
The standard Ritter scheme~\citep{Ritter88} can be applied only if 
the pressure scale height at the stellar surface is much smaller than 
the stellar radius ($H_P \ll R$). 

We do not consider rotation in this study, but assume  non-conservative mass
transfer as predicted by previous binary models including the effects of
rotation~\citep{Petrovic05, Yoon10, Langer12}.  In these studies, the mass
accretor is quickly spun up to critical rotation during a mass transfer
phase, and the stellar wind mass loss rate from the accretor dramatically
increases accordingly, which effectively results in highly non-conservative
mass transfer.  In our calculation, therefore, we assume the ratio of the
accreted mass to the transferred mass (i.e., the mass accretion efficiency
$\beta$) is 0.2.  In this picture, the non-accreted matter is blown away from
the binary system as a fast wind from the mass accretor, and the corresponding
angular momentum loss from the orbit is calculated accordingly.

We considered four different
primary star masses and two different metallicities: $M_1=$ 11, 13, 16, and
18~$\mathrm{M_\odot}$, and $Z =$ 0.02 and 0.007, which roughly correspond to
solar and Large Magellanic Cloud (LMC) metallicities respectively.   The mass
ratio is fixed to $q = 0.9$ for all models in our grid.  To make a robust
prediction on the event rates of SNe Ib and IIb from binary systems, we would need to
consider a wider range of mass ratio~\citep{Claeys11}, but it is beyond the
scope of this paper.  

\begin{figure}
\begin{center}
\includegraphics[width = 0.98\columnwidth]{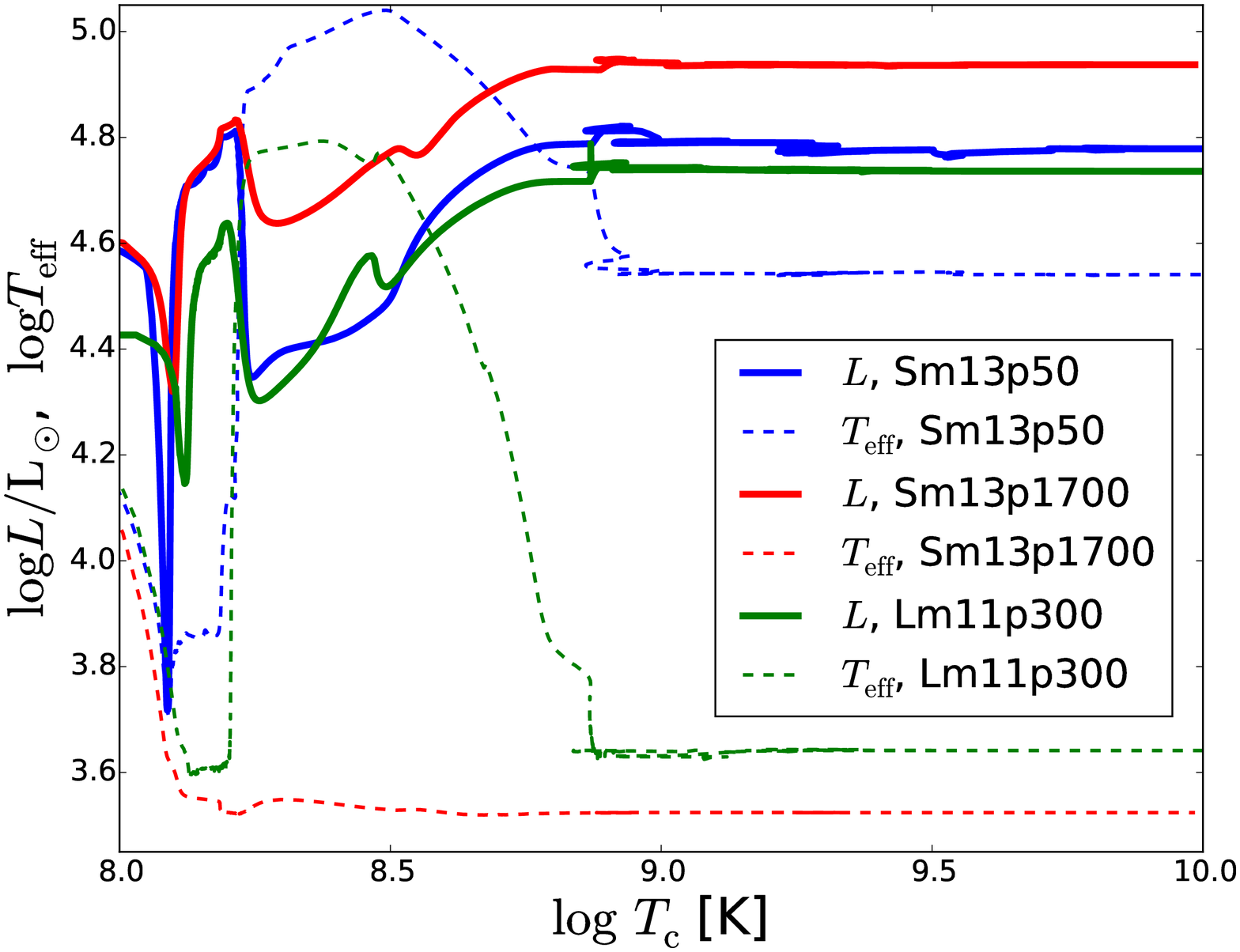}
\caption{Evolution of the luminosity (solid line) and surface temperature (dashed line)
of the primary star in Sm13p50 (blue), Sm13p1700 (red) and Lm11p300 (green), 
as a function of temperature at the center. The luminosity and surface temperature
remain almost constant after $T_\mathrm{c}$ increases beyond $10^9$~K.  
}\label{fig:change}
\end{center}
\end{figure}

The properties of the primary star at the end of the calculation for each
sequence are summarized in Tables~\ref{tab1},~\ref{tab2}, and~\ref{tab3}.  LMC
and solar metallicity models have reference names starting with L and S,
respectively (Lm11p100, Sm13p500, etc.). Names starting with T refer to
test models with a reduced mass loss rate at solar metallicity (e.g., Tm11p20; see Section~\ref{sect:massloss}).
Each name also contains the information of the initial mass of the primary
star and the initial period.  For example, Lm11p100 refers to the sequence with
$M_\mathrm{1, init} = 11~\mathrm{M_\odot}$ and $P_\mathrm{init} = 100$~d at LMC
metallicity. 

The model evolution is halted when the central temperature $T_\mathrm{c}$
reaches $10^9$~K in most cases.  The evolution of Lm11p300, Sm13p50 and Sm13p1700
is calculated to the pre-SN stage to check if there would be any
significant change in the surface properties (i.e., $L$ and $T_\mathrm{eff}$)
from this point.  As shown in Fig.~\ref{fig:change}, the envelope structure
remains almost unchanged, because of very short evolutionary time from
$T_\mathrm{c} = 10^9$~K until core-collapse ($\sim 10$~yr).  In the sequences
Sm10p50 and Sm10p200, the carbon-oxygen core becomes electron-degenerate as its
mass approaches 1.4~$\mathrm{M_\odot}$.  Off-center oxygen burning starts in
these models before reaching $T_\mathrm{c} = 10^9$~K, and we stopped the
calculation at this point because the time step becomes extremely small. It is
likely that these models result in SN Ib progenitors, but we leave their final
fate open in the present study.

\section{Results}\label{sect:results}

\begin{figure*}
\begin{center}
\includegraphics[width = 0.5\textwidth, angle=-90]{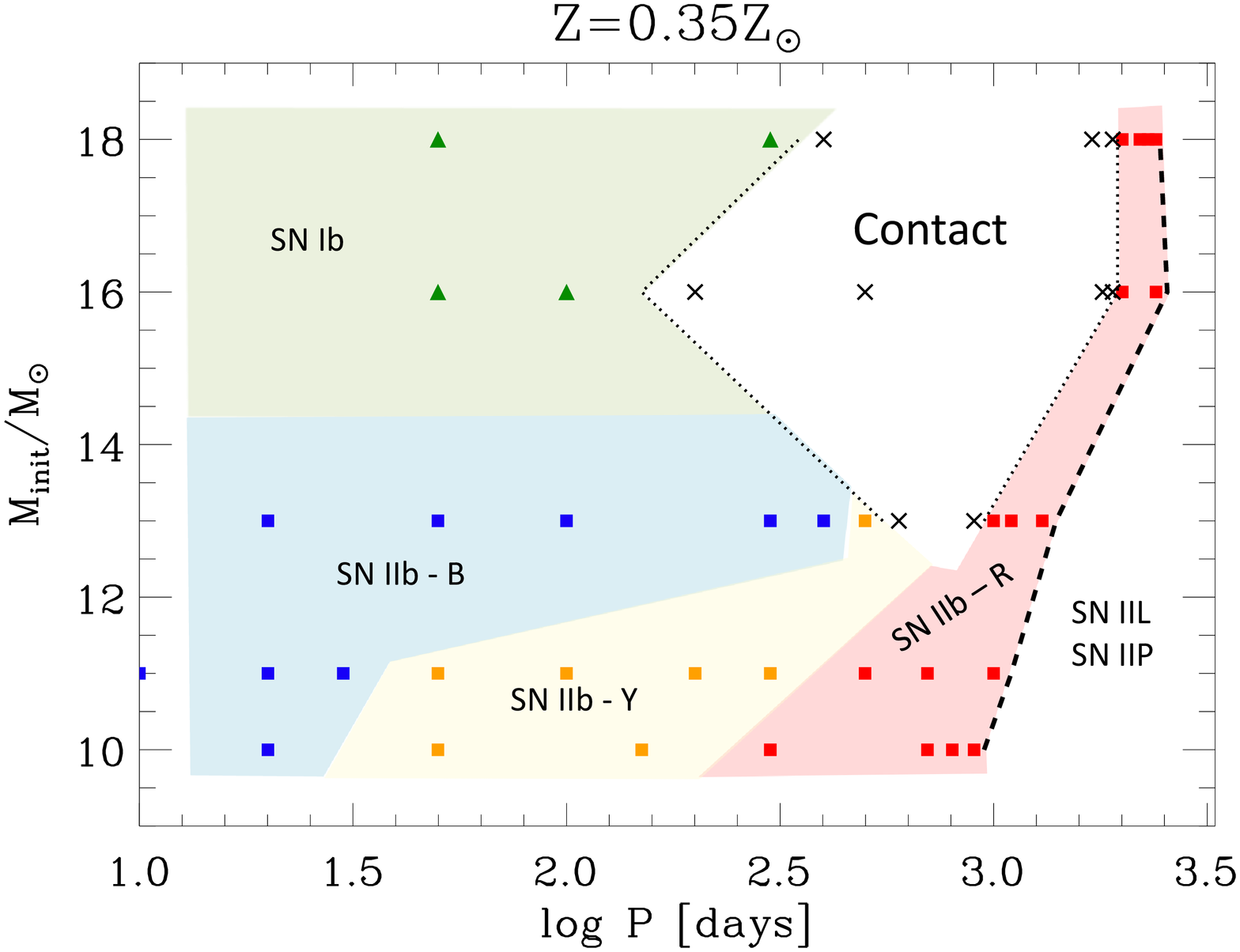}
\vskip   0.5cm
\includegraphics[width = 0.5\textwidth, angle=-90]{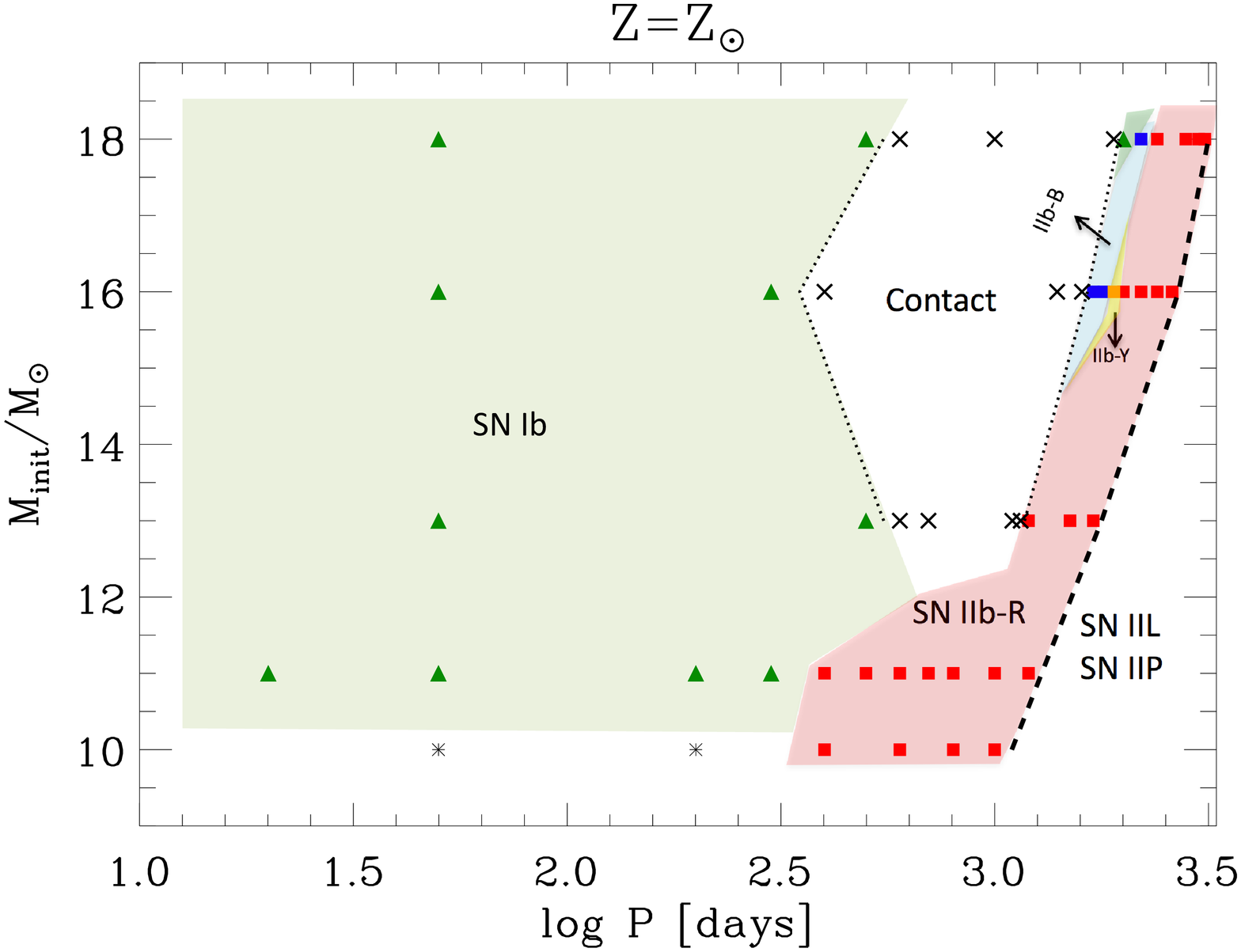}
\caption{Grid points of the calculated binary sequences for LMC (upper panel) and solar (lower panel) metallicities 
on the plane spanned by the initial orbital period and the initial mass of the primary star. 
The predicted final fates of the primary stars are marked by different symbols.  
Triangle and square symbols denote SN Ib and SN IIb, respectively. Blue, orange and red colors for the square symbol
indicate SN IIb progenitors of blue (SN IIb-B), yellow (SN IIb-Y) and red (SN IIb-R) colors, respectively (see the text). 
Sequences where the binary system becomes a contact binary are marked by a black cross. The black star symbols correspond
to the models whose evolution was stopped before reaching $T_\mathrm{c} = 10^9$~K. 
}\label{fig:result}
\end{center}
\end{figure*}

Within the parameter space we considered in the present study, 
the primary star undergoes Case B mass transfer: Roche-lobe overflow after core hydrogen exhaustion 
and before core helium exhaustion. 
For our discussion, we further distinguish Case B as follows:
\begin{itemize}
\item Case EB (early Case B):  mass transfer while the primary star crosses the Hertzsprung gap on the HR diagram.  
\item Case LB (late Case B): mass transfer after the hydrogen envelope of the primary star becomes fully convective. 
\end{itemize}
After Case EB/LB mass transfer, a second mass transfer from the primary star may occur after core helium exhaustion.
This case is referred to as Case EBB/LBB.

We can roughly predict the resultant SN type from our models based on their
structures.  A SN IIb progenitor must contain some hydrogen in the envelope but
the hydrogen envelope mass ($M_\mathrm{H, env}$) should not exceed a certain
limit as otherwise the resultant supernova would appear as a SN IIL or IIP.
This upper limit of $M_\mathrm{H, env}$  would depend on the supernova
parameters including supernova energy, total ejecta mass, and chemical composition. 
In the present study, we assume the condition of $M_\mathrm{H, env} \le
1.0~\mathrm{M_\odot}$  for SN IIb progenitors. 

In addition, we categorize our SN IIb progenitor models into three groups
according to their surface temperatures at the final stage as the following: 
\begin{itemize}
\item SN IIb-B: blue progenitors ($\log T_\mathrm{eff}/\mathrm{K} > 3.875$; e.g. SN 2008ax and SN 2016gkg)
\footnote{Some of SN IIb-B models in our grid are too compact ($R < \sim 10~\mathrm{R_\odot}$) to be called a giant star, 
and therefore we do not refer our blue progenitor models to blue supergiant progenitors.}
\item SN IIb-Y: yellow supergiant (YSG) progenitors ($3.681 \le \log T_\mathrm{eff}/\mathrm{K} \le 3.875$; e.g., SN 2011dh)
\item SN IIb-R: red supergiant (RSG) progenitors ($\log T_\mathrm{eff}/\mathrm{K} < 3.681$; e.g., SN 1993J and SN 2013df)
\end{itemize}
Our temperature criterion for YSGs is adopted from~\citet{Drout09}.  Note that
in the literature, there is ambiguity about the definition of YSGs: some
authors consider progenitors of SN 1993J and 2013df as
YSGs~\citep[e.g.,][]{Bersten12}, which are RSGs according to our definition.
In the case where no hydrogen is left, we assume that a SN Ib would occur.  The
remaining helium mass in this case is higher than about
$1.5~\mathrm{M_\odot}$, for which helium cannot be easily hidden in supernova
spectra for the final mass range of our SN Ib models
($3.0~\mathrm{M_\odot}\lesssim M \lesssim 6.6~\mathrm{M_\odot}$;
\citealt{Hachinger12, Dessart15}). 

We summarize the result of our calculations in Fig.~\ref{fig:result}, 
where  the predicted final fate of the primary star is indicated
in the plane spanned by the initial period and initial mass. 
As shown in the figure, SN IIb progenitors are expected for a larger
parameter space at LMC metallicity than at solar metallicity.  The
figure also shows the dependence of the progenitor type on the initial orbital
period ($P_\mathrm{init}$) and the initial mass of the primary star
($M_\mathrm{init}$), as explained below in greater detail.

\subsection{LMC metallicity models}

\begin{figure*}
\begin{center}
\includegraphics[width =0.98\columnwidth]{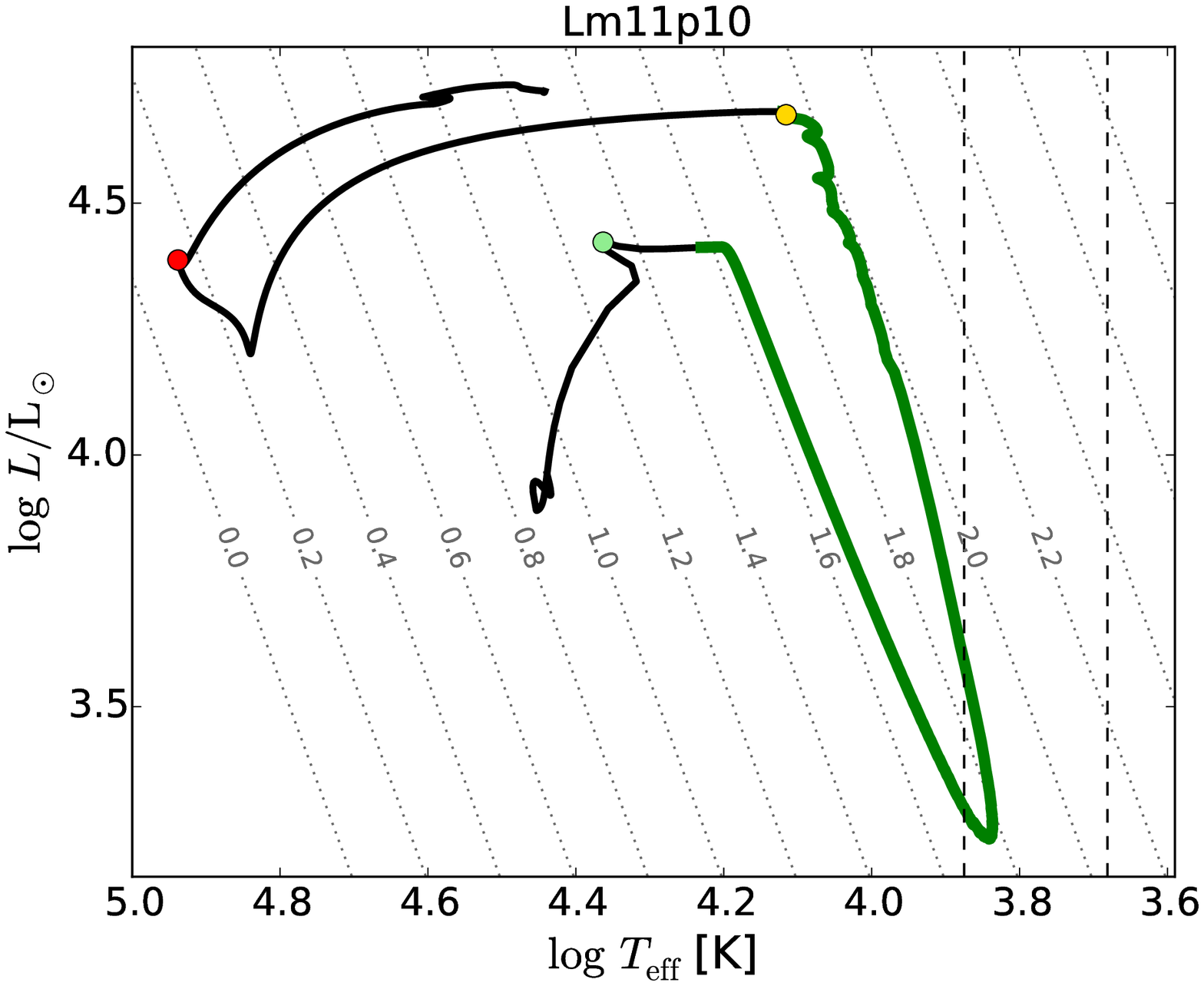}
\includegraphics[width =0.98\columnwidth]{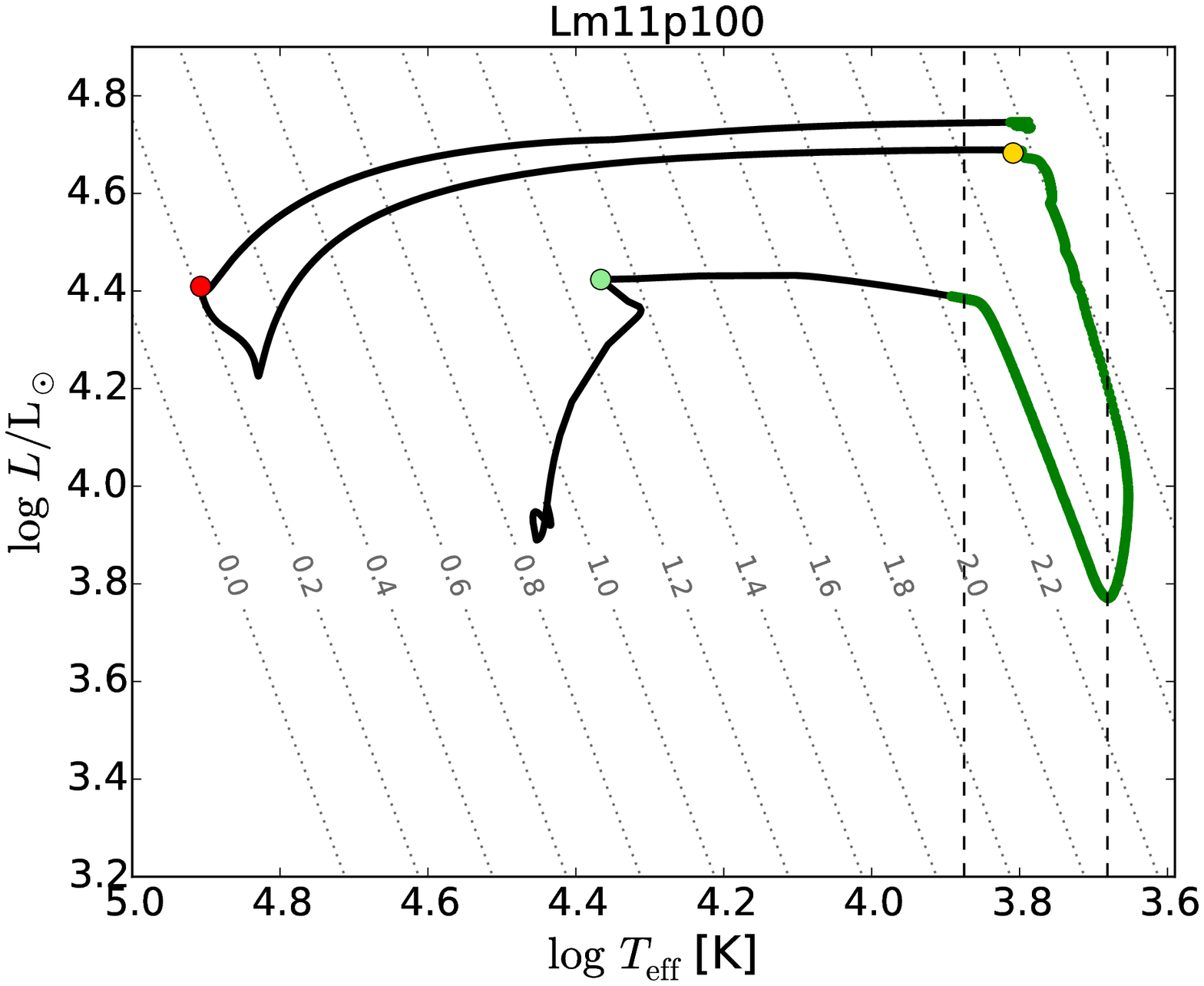}
\includegraphics[width =0.98\columnwidth]{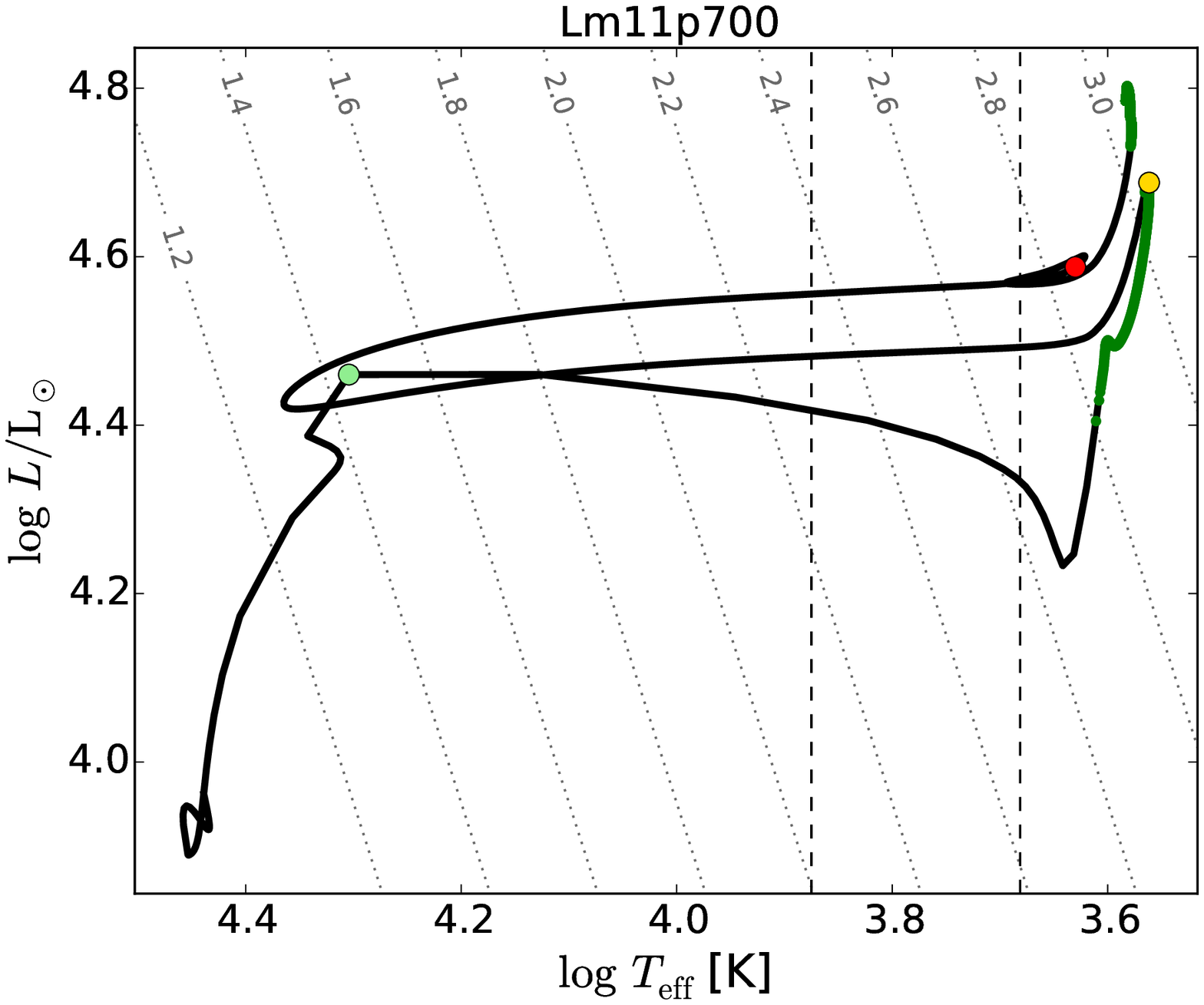}
\includegraphics[width =0.98\columnwidth]{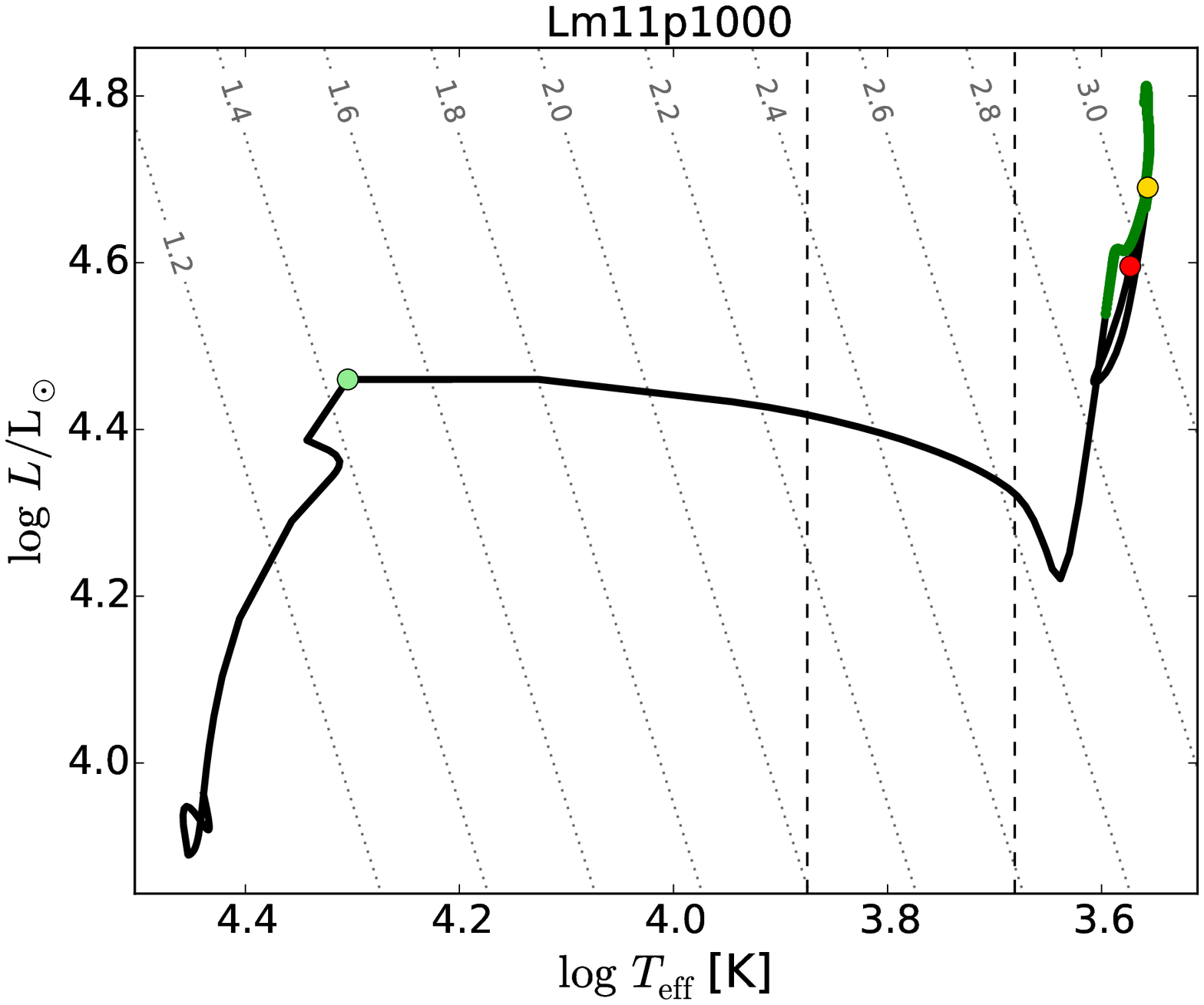}
\caption{
HR diagrams for systems with a primary star of  $M_\mathrm{init} = 11~\mathrm{M_\odot}$ at LMC metallicity. 
The secondary mass is 9.9~$\mathrm{M_\odot}$. The adopted mass accretion efficiency is 0.2.
The mass transfer phases are marked by green color on each evolutionary track.  
The initial periods of the systems are 10 (top left), 100 (top right), 700 (bottom left), and 1000 (bottom right)~days, as indicated 
on top of each figure. The labels with dotted contour lines indicate $\log R/\mathrm{R_\odot}$. 
The vertical dashed lines give the boundary values for yellow supergiants ($3.681 \le \log T_\mathrm{eff}/\mathrm{K} \le 3.875$). 
The light green, yellow and red circles denote core hydrogen exhaustion, onset of core helium burning and core helium exhaustion, respectively. 
}\label{fig:hr11lmc}
\end{center}
\end{figure*}

It is well known that an initially tighter orbit leads to a deeper stripping of
the hydrogen envelope via mass transfer.  This effect of the initial separation
can be observed in Fig.~\ref{fig:hr11lmc}, as an example with $M_\mathrm{init}
= 11~\mathrm{M_\odot}$.  

With relatively short initial orbits ($P_\mathrm{init} =$ 10~d and 100~d), Case
EB mass transfer occurs.  In this case, most of the hydrogen envelope remains
radiative at the onset of mass transfer, and strong mass loss leads to thermal
disequilibrium of the envelope.  The luminosity of the primary star rapidly
drops accordingly~\citep[e.g., see][]{DeLoore92, Eggleton11}
until the total mass decreases  to about 7~$\mathrm{M_\odot}$
and 8~$\mathrm{M_\odot}$ for $P_\mathrm{init} =$~10~d and 100~d, respectively.
The star gradually becomes brighter thereafter as the surface mass fraction of
hydrogen decreases.  Once the mass transfer stops, both the radius and the
luminosity of the star decrease until thermal equilibrium is
restored. 

After core helium
exhaustion, the envelope expands again until the end point.  The surface
temperature at the end point is $\log T_\mathrm{eff}/\mathrm{K} = 4.44$ and 3.79 for
$P_\mathrm{init} =$~10 and 100~d, respectively. As discussed in
Section~\ref{sect:mhenv} below in more detail, the final surface temperature
tends to be lower for a higher hydrogen envelope mass that remains until the
end, which is in turn determined by the initial orbital separation. 

In the sequences with  $P_\mathrm{init} =$700~d and 1000~d, mass transfer
occurs when the hydrogen envelope of the primary star is fully convective (Case
LB). The response to mass loss of a convective envelope is very different from
that of a radiative envelope~\citep{DeLoore92, Eggleton11}. As shown in
Fig.~\ref{fig:hr11lmc}, the radius initially increases once the primary star
begins to transfer mass to its companion.  In the sequence of $P_\mathrm{init}
=$~700~d, after the mass transfer stops, the star undergoes a blueward evolution
on the HR diagram.  The star turns to the red again and becomes a RSG at the final
stage.  On the other hand, the HR diagram of the sequence with $P_\mathrm{init}
=$~1000~d does not show the blue loop.  This is because the hydrogen envelope
mass remains sufficiently large to preserve the RSG structure.   

\begin{figure}
\begin{center}
\includegraphics[width =0.95\columnwidth]{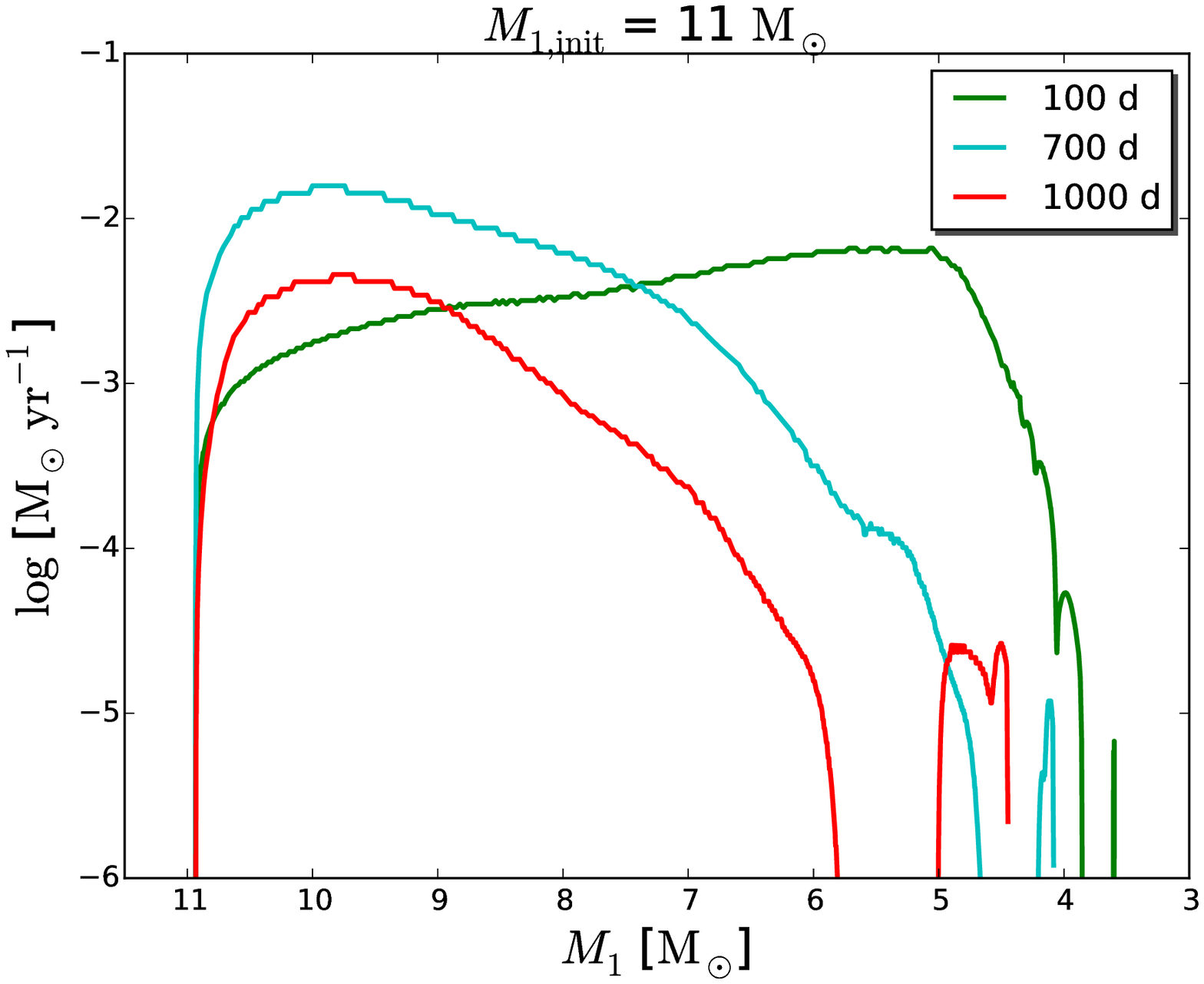}
\includegraphics[width =0.95\columnwidth]{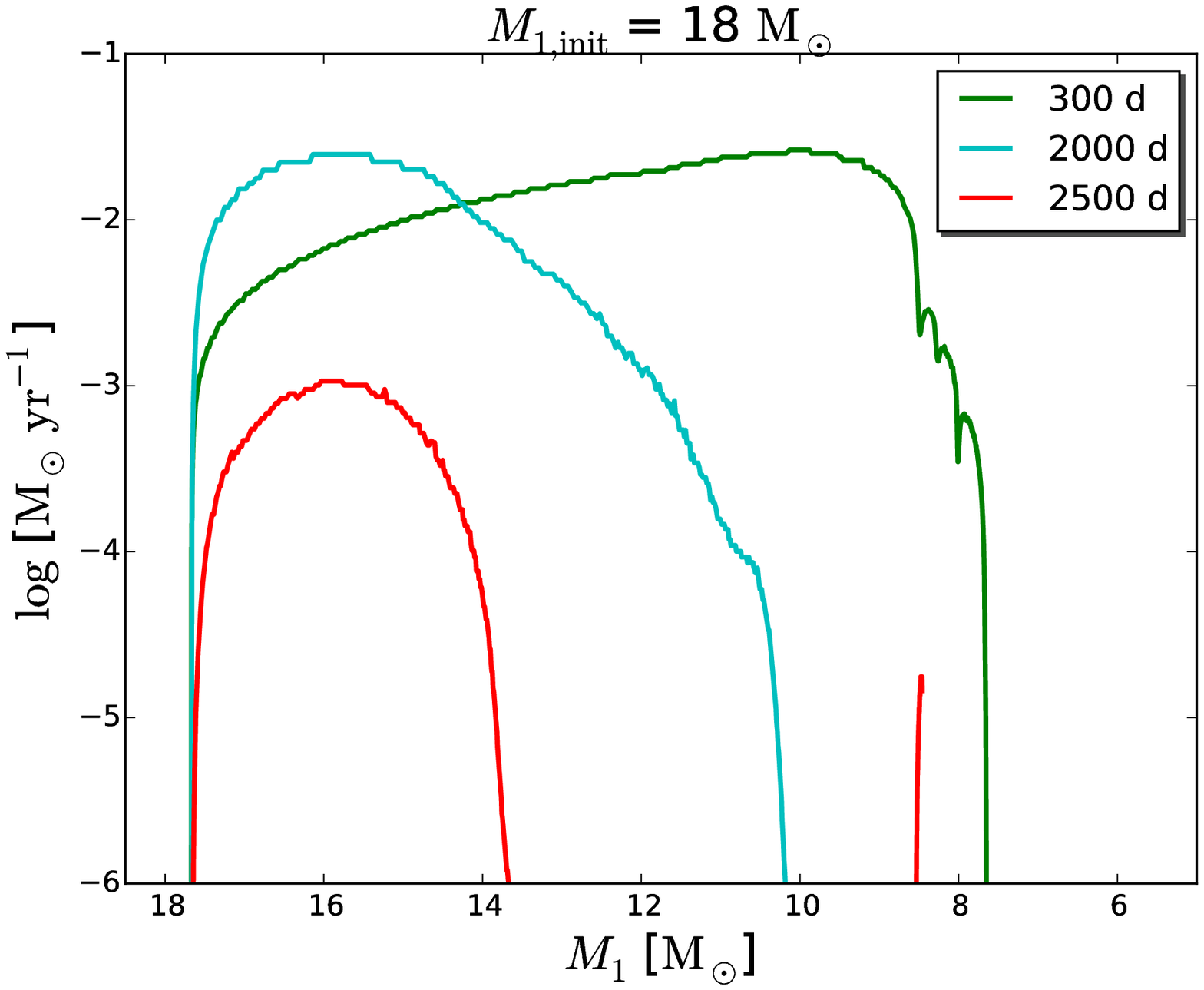}
\caption{The mass transfer rates as a function of the total mass of the primary star, for systems
with $M_\mathrm{1, init} = 11~~\mathrm{M_\odot}$ (upper panel) and  $M_\mathrm{1, init} = 18~~\mathrm{M_\odot}$ 
(lower panel) at LMC metallicity. The initial periods are indicated by different colors. 
}\label{fig:mdotlmc}
\end{center}
\end{figure}

In Fig.~\ref{fig:mdotlmc}, the evolution of the mass transfer rate from the
11~$\mathrm{M_\odot}$ and 18~$\mathrm{M_\odot}$ primary stars is presented for
three different initial periods. We note that  Case EB mass transfer
($P_\mathrm{init} =$~100~d for 11~$\mathrm{M_\odot}$ and $P_\mathrm{init}
=$~300~d for 18~$\mathrm{M_\odot}$) lasts longer than Case LB mass transfer,
which explains the deeper stripping of the hydrogen envelope with  Case EB. With
Case LB mass transfer, the duration and mass transfer rate become smaller with a
longer initial period.  There are two mass transfer phases for the
11~$\mathrm{M_\odot}$ model.  The
first one is responsible for most of
the stripping of the hydrogen envelope.  The second one is a brief phase
with a lower mass transfer rate, which occurs after core helium exhaustion as
the primary star tends to expand during the final evolutionary stages.  

This second mass transfer is less favored in more massive systems
(Fig.~\ref{fig:mdotlmc}; see also Tables~\ref{tab1},~\ref{tab2},
and~\ref{tab3}).  This is because of the following two factors. 1) The Case EBB
mass transfer (i.e., the second mass transfer after the Case EB mass transfer)
is commonly observed for relatively low-mass stars~\citep[e.g.,][]{Yoon10}.
The envelope expansion after core helium exhaustion is more
prominent with a more compact stellar core.  On the other hand, higher mass
stars ($M_\mathrm{init} \gtrsim 13~\mathrm{M_\odot}$) tend to remain more
compact provided $M_\mathrm{H,env} <
0.15~\mathrm{M_\odot}$ after the Case EB mass transfer.  2) The Case LBB mass
transfer (i.e., the second mass transfer after the Case LB mass transfer) is
also driven by the expansion of the remaining hydrogen envelope.  For more
massive systems, however, stronger stellar winds tend to make the orbit
wider, preventing the primary star from undergoing further mass transfer. 

Although mass transfer is the main determining factor for the final envelope
structure of SN progenitors in binary systems, the effect of wind mass loss
can also play an important role.  As discussed by \citet{Yoon10},  mass
transfer cannot completely remove hydrogen from the primary star even with
Case EB.  Whether the progenitor can retain some hydrogen until core collapse
depends on the subsequent history of mass loss by stellar winds.  Stellar winds
are stronger for higher masses, and no hydrogen is left in Case EB systems with
$M_\mathrm{init} \ge 16~\mathrm{M_\odot}$ at $Z = Z_\mathrm{LMC}$
(Fig.~\ref{fig:result}; Table~\ref{tab1}).  

With our assumption of $\beta = 0.2$, the Case EB
and Case LB regimes are separated by the contact regime  for the sequences with
$M\ge13~\mathrm{M_\odot}$ (Fig.~\ref{fig:result}). With a larger value of
$\beta$, the regime for contact would be enlarged.  

In terms of the final fate (Fig.~\ref{fig:result}), solutions for SN IIb-B and
SN IIb-Y progenitors are found only for $M\le13~\mathrm{M_\odot}$ in our
considered parameter space.  SN IIb-R solutions are found for all the considered
mass range.  The tendency that the surface temperature gradually decreases with increasing
$P_\mathrm{init}$ for a given initial mass can be explained by the fact that
less hydrogen is removed in the case of a wider orbit. 

\subsection{Solar metallicity models}

The mass transfer history of solar metallicity models is qualitatively similar
to that of LMC metallicity models. A shorter initial orbit leads to more
significant stripping of the hydrogen envelope, and vice versa.  However, the
role of stellar winds becomes more important, compared to the LMC case. 

The most notable difference between solar and LMC metallicity models is
observed in the outcome of Case EB mass transfer.  As shown in
Fig.~\ref{fig:result}, hydrogen is completely removed via stellar winds during
the post Case EB mass transfer phase (see also Table~\ref{tab2}). The solutions
of SN IIb-B and SN IIb-Y that are found with LMC metallicity with $M \le
13~\mathrm{M_\odot}$ are replaced by SN Ib solutions. 

Strong winds also significantly affect the evolution with Case LB mass
transfer.  Narrow regimes for SN IIb-B solutions  appear with $M =
16~\mathrm{M_\odot}$ and 18~$\mathrm{M_\odot}$, which are not found at LMC
metallicity. This is because the primary star loses a significant amount of
hydrogen during the blue loop phase where the WR mass loss rate of
\citet{Nugis00} is applied (see Section~\ref{sect:massloss} below) after Case LB
mass transfer  (Fig.~\ref{fig:hr16zsol}). At $M_\mathrm{init} =
 16~\mathrm{M_\odot}$, a SN IIb-Y solution is also found in-between SN IIb-B and
SN IIb-R regimes (Figs.~\ref{fig:result} and~\ref{fig:hr16zsol}). 

In general, RSG solutions are dominant for SN IIb progenitors at solar
metallicity. However, many of the SN IIb-R progenitor models have higher surface
temperatures ($\log T_\mathrm{eff}/\mathrm{K} > 3.6$) than those of ordinary RSGs observed
in the local universe (see Table~\ref{tab2}).  This  is  consistent with
the progenitor properties of SN 1993J and SN 2013df, for which $\log
T_\mathrm{eff}/\mathrm{K} \simeq 3.63$ was inferred observationally~\citep{Maund04,
VanDyk14}.

\begin{figure}
\begin{center}
\includegraphics[width =1.0\columnwidth]{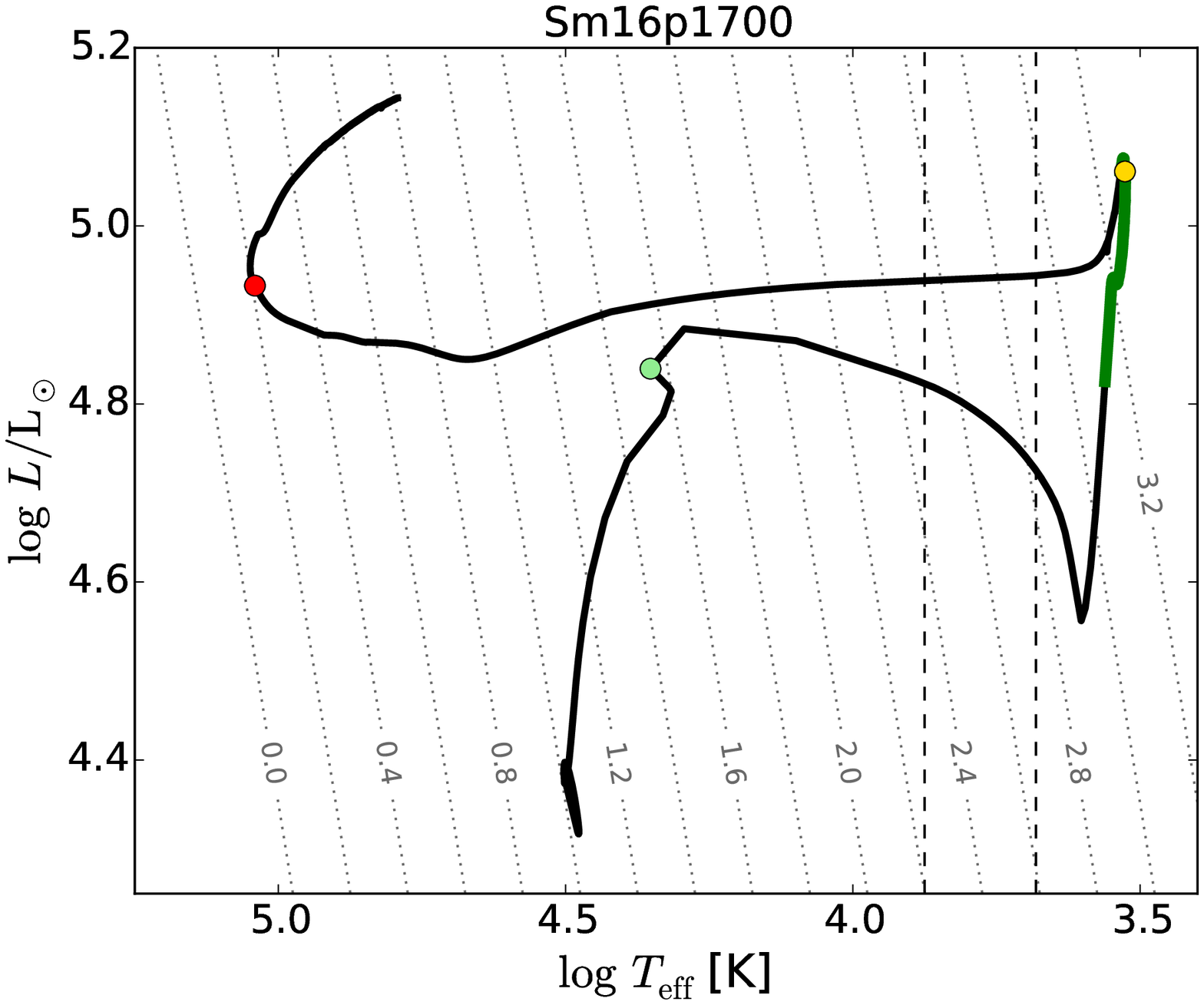}
\includegraphics[width= 1.0\columnwidth]{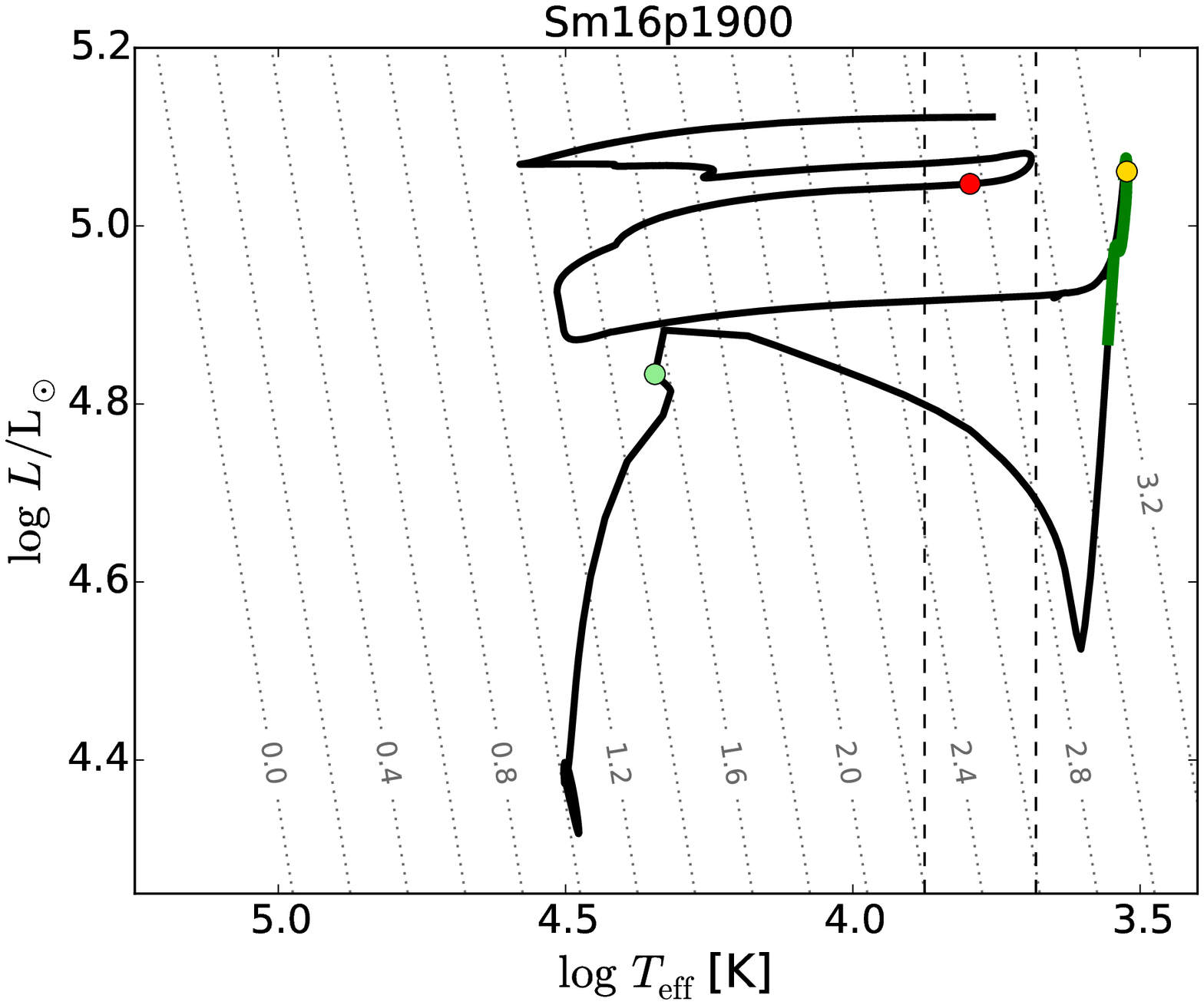}
\includegraphics[width= 1.0\columnwidth]{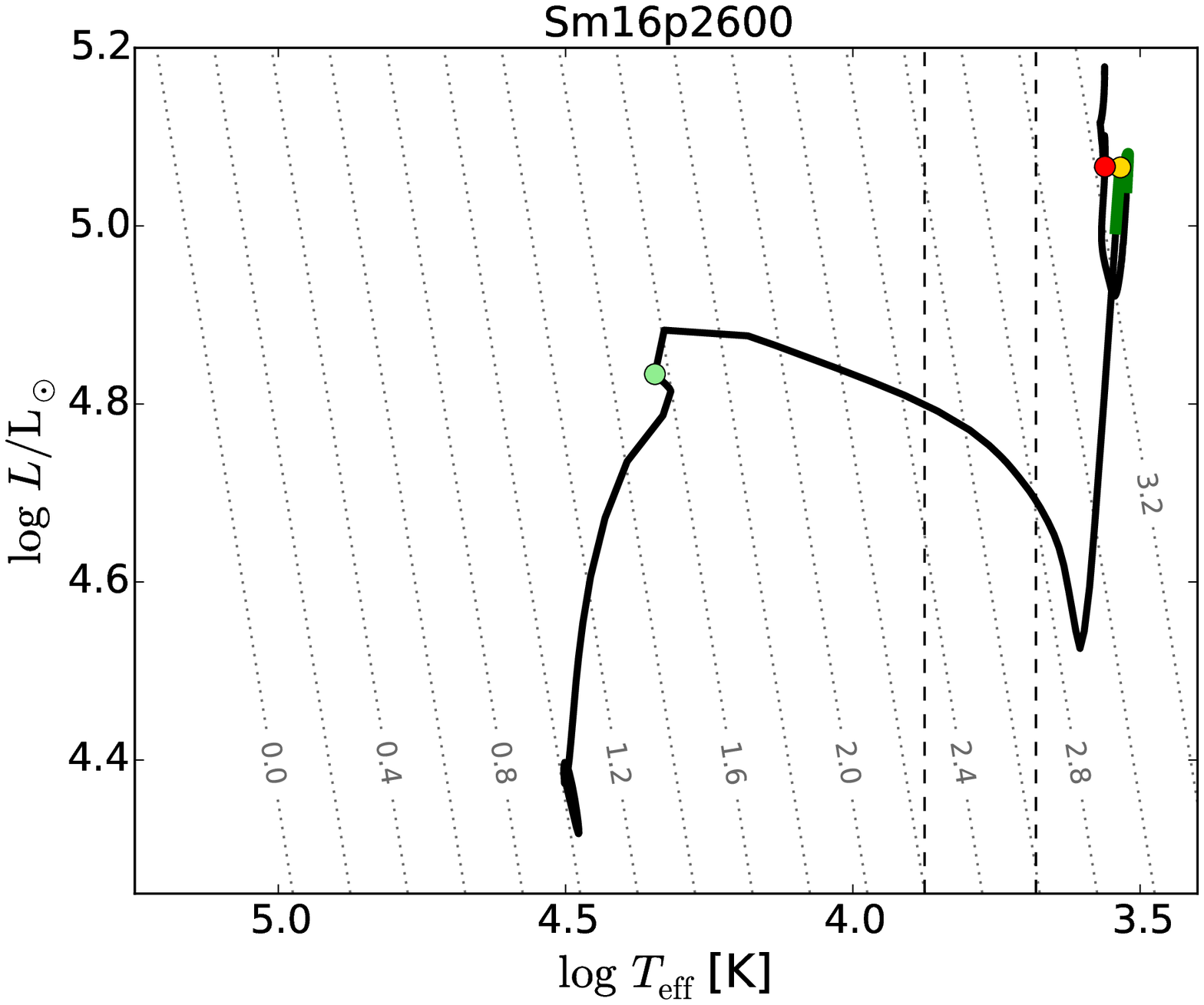}
\caption{
HR diagrams for systems with a primary star of $M_\mathrm{init} = 16~\mathrm{M_\odot}$ at solar metallicity. 
The secondary mass is 14.4~$\mathrm{M_\odot}$. 
The initial periods of the systems are 1700 (top), 1900 (middle), and 2600 (bottom)~days. 
The mass transfer phases are marked by green color on each evolutionary track.  
The labels with dotted contour lines indicate $\log R/\mathrm{R_\odot}$. 
The vertical dashed lines give the boundary values for yellow supergiants ($3.681 \le \log T_\mathrm{eff}/\mathrm{K} \le 3.875$). 
The light green, yellow and red circles denote core hydrogen exhaustion, onset of core helium burning and core helium exhaustion, respectively. 
}\label{fig:hr16zsol}
\end{center}
\end{figure}

\subsection{Effect of WR winds}\label{sect:massloss}

As discussed above, the evolution of the primary star after the major mass
transfer phase is critically affected by stellar wind mass loss. In particular,
the primary star spends most of the remaining evolutionary time on the helium
main sequence as a hydrogen-poor star after Case EB mass transfer.  The WR mass
loss rate by \citet{Nugis00} (hereafter, the NL rate) is applied for this case.
However, the NL rate was inferred from Galactic WR stars, which are much more
massive ($M \gtrsim 10~\mathrm{M_\odot}$) and therefore much more luminous than
our primary star models during the post mass transfer phase ($
3~\mathrm{M_\odot} \lesssim M \lesssim 7~\mathrm{M_\odot}$).  The validity of
the NL rate for these relatively low-mass hydrogen-poor hot stars has hardly
been investigated previously (see the discussion by \citealt{Yoon15}).  

For their SN Ib progenitor models in binary systems, \citet{Yoon10} have applied
a lower mass loss rate than the NL rate, if $\log
L/\mathrm{L}_\odot < 4.5$ during the post mass transfer phase.  As a
result, a small amount of hydrogen is left in many of their SN Ib progenitor
models, which can be considered as SN IIb-B progenitors according to our
definition. 

To investigate further the effect of wind mass loss, we made a model grid with
a reduced mass loss rate using the scaling factor $\eta$ = 0.48 on 
the Dutch mass loss rate recipe in the MESA code,
at solar metallicity (Table~\ref{tab3}). The mass loss rate with this reduction
factor becomes comparable to that of LMC metallicity with the fiducial value of
$\eta = 1.0$.  As shown in Table~\ref{tab3}, this reduction of the mass loss
rate gives a result similar to that of the LMC metallicity models.  Solutions
for SN IIb progenitors are found with the sequences with $M_\mathrm{init} =
11~\mathrm{M_\odot}$ and 13~$\mathrm{M_\odot}$ for $P_\mathrm{init} \le 400$~d,
for which SN Ib solutions are obtained with our fiducial mass loss rate.  With
$M_\mathrm{init} = 11~\mathrm{M_\odot}$, solutions for SN IIb-B, SN IIb-Y, and
SN IIb-R can be found in the order of increasing initial periods. With
$M_\mathrm{init} = 13~\mathrm{M_\odot}$, only SN IIb-B is predicted. This
result is also consistent with those of \citet{Benvenuto13} and
\citet{Folatelli15}, who found a SN IIb-Y solution for SN 2011dh and a SN
IIb-B solution for SN 2008ax, respectively, at solar metallicity.  Apparently,
these authors did not apply WR mass loss in their binary models. 

This consideration should be taken as a caveat when we discuss the metallicity
dependence of the predicted SN type (see Section~\ref{sect:metallicity})


\section{Properties of SN II\lowercase{b} progenitors}\label{sect:snIIb}

\subsection{Final mass - luminosity relation}

\begin{figure}
\begin{center}
\includegraphics[width =1.0\columnwidth]{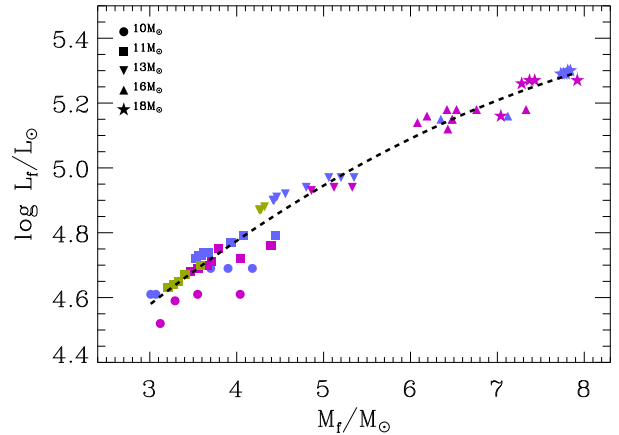}
\caption{The final mass-luminosity relation of SN IIb progenitors. 
The initial primary mass is marked by different symbols: circle, square, inverted triangle, triange, and star
denote 10, 11, 13, 16 and 18~$\mathrm{M_\odot}$, respectively. 
Purple and blue indicate
solar and LMC metallicities, respectively. Green gives the result with 
a reduced mass loss rate (i.e., $\eta = 0.48$) at solar metallicity.  
The dashed line is the best fit of the data with a second order polynomial. 
}\label{fig:mfL}
\end{center}
\end{figure}

A few SN IIb progenitors have been directly identified in pre-SN images,
including those of SN 1993J, SN 2008ax, SN 2011dh, SN 2013bf and SN
2016gkg~\citep{Aldering94, Maund04, Maund11, VanDyk11, VanDyk14, Kilpatrick16,
Tartaglia16}.  Their properties are summarized in Table~\ref{tab4}.  In these
studies, the progenitor masses at the pre-SN stage were inferred from their
bolometric luminosities. 

In Fig.~\ref{fig:mfL}, we present the final mass-luminosity relation of our SN IIb
progenitor models. The luminosity at the pre-SN stage correlates with
the initial and final masses.   For a given final luminosity, however,
the scatter of the final mass can be as large as 1.0~$\mathrm{M_\odot}$. The
bolometric luminosity at the pre-SN stage is mainly determined by helium shell
burning, which is in turn largely determined by the CO core mass. As shown in
Tables~\ref{tab1},~\ref{tab2} and~\ref{tab3}, different final CO core masses
may result from different initial orbits for a given initial mass. This explains
the scatter of the final masses for a given luminosity. Likewise,
for a given initial mass, the final luminosity may vary by up to 0.2~dex in
terms of  $\log L/\mathrm{L_\odot}$. 

The final mass-luminosity relation shown in  Fig.~\ref{fig:mfL} roughly gives $M_f
\simeq$ 6.0, 5.0, and 4.9~$\mathrm{M_\odot}$ for SN 1993J, SN 2011dh and SN
2013df, respectively, for which the progenitor luminosity is fairly well
constrained.  The corresponding initial masses are about 16~$\mathrm{M_\odot}$
for SN 1993J and 13~$\mathrm{M_\odot}$ for SN 2011dh and SN 2013df, in
agreement with previous work.

\subsection{Final position on the HR diagram and optical brightness}\label{sect:finalHR}

\begin{figure}
\begin{center}
\includegraphics[width =1.0\columnwidth]{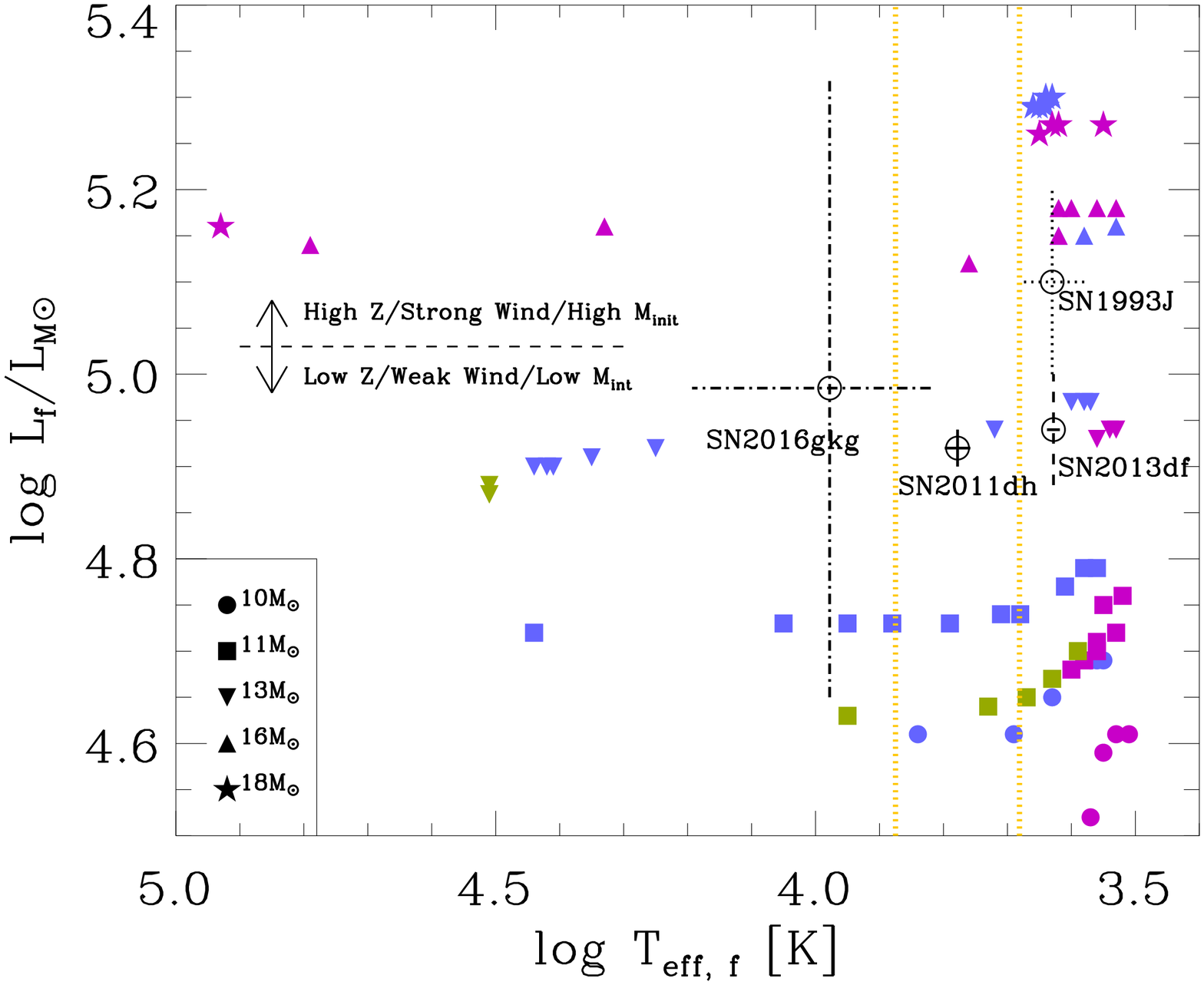}
\includegraphics[width =1.0\columnwidth]{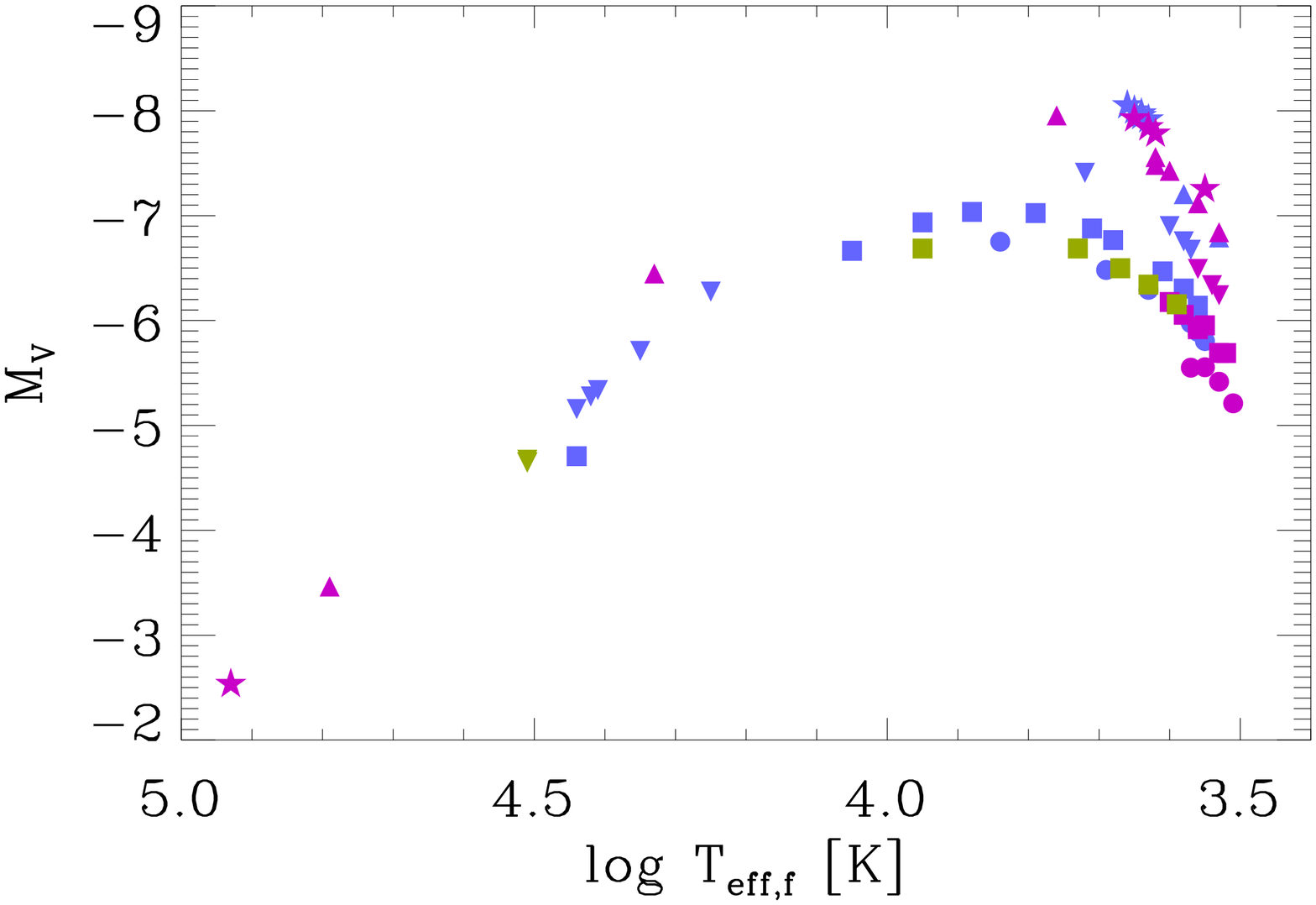}
\caption{\emph{Upper panel:} The final positions of our SN IIb models on the HR diagram. 
The initial primary mass is marked by different symbols: circle, square, inverted triangle, triangle, and star
denote 10, 11, 13, 16 and 18~$\mathrm{M_\odot}$, respectively. 
Purple and blue indicate
solar and LMC metallicities, respectively. Green gives the result with 
a reduced mass loss rate (i.e., $\eta = 0.48$) at solar metallicity.  
The orange dotted lines mark the boundaries for yellow supergiants: 
$3.681 \le \log T_\mathrm{eff}/\mathrm{K} \le 3.875$.  
The black dashed line roughly gives the dividing line of SN IIb-B solutions
for different metallicities, mass loss rates and/or initial masses (see the text
for details). The open circles mark the positions of 
four observed SN IIb progenitors (SN 1993J, SN 2011dh, SN 2013df, and SN 2016gkg; see Table~\ref{tab4}). 
\emph{Lower panel:} Corresponding V-band magnitudes with black-body approximation. 
}\label{fig:hr}
\end{center}
\end{figure}

The positions on the HR diagram of our SN IIb progenitor models are shown in
Fig.~\ref{fig:hr}. Given the limited number of our models, this figure cannot
properly reflect the real population of SN IIb progenitors.  However, a few
important features can be found here. 

Note that RSG progenitors (SN IIb-R) like SN 1993J and SN 2013bf can exist for
a wide range of luminosities (i.e, for a wide range of initial masses)
regardless of metallicity, and can be relatively easily explained by Case LB
mass transfer.  

On the other hand,  SN IIb-B and SN IIb-Y progenitors would be preferentially
produced from relatively low-mass progenitors,  as also implied by
Fig.~\ref{fig:result}.  For $\log L_f/\mathrm{L_\odot} < 5.0$, which corresponds to
$M_\mathrm{1,init} \lesssim 13~\mathrm{M_\odot}$ solutions for SN IIb-B and
SN IIb-Y can be obtained only with LMC metallicity or with a reduced mass loss
rate ($\eta = 0.48$) at solar metallicity. These solutions are obtained mostly
with Case EB mass transfer.  SN IIb-B and IIb-Y solutions  for  $\log
L_f/\mathrm{L_\odot} > 5.0$ ($M_\mathrm{1,init} \gtrsim 16~\mathrm{M_\odot}$)
result from strong mass loss after Case LB mass transfer, and are rarely
achieved with $\eta = 1.0$ at solar metallicity, as implied by
Fig.~\ref{fig:result}. 

The host galaxies of SN 2011dh and SN 2016gkg (M51 and NGC 613,
respectively), for which the progenitors belong to SN IIb-B or SN IIb-Y type,
are not particularly metal poor. This might imply a lower mass loss rate from the
progenitors than the fiducial mass loss rate adopted in our
study~\citep[cf.][]{Benvenuto13}.  However, the parameter space of our models
is limited to stable mass transfer systems.  In reality, multiple binary
channels may exist (e.g., common envelope ejection) for SN IIb-B/IIb-Y
progenitors, and this issue deserves a more detailed future study.  

Fig.~\ref{fig:hr} also shows the V-band magnitudes of our SN IIb progenitor
models, which were calculated with the black-body approximation.   In general, our
models predict that SN IIb progenitors are much brighter ( $M_\mathrm{V} < -6.0$ for
most cases)  than  hydrogen-free SN Ib/Ic progenitors ($M_\mathrm{V} > -4.5$;
\citealt{Yoon12};  see Section~\ref{sect:snIb} below) in the optical. Therefore, it is not
surprising that  SN IIb progenitors have be more commonly identified
than SN Ib/Ic progenitors so far~\citep[see][for a recent review]{Smartt15},
despite the fact that the SN Ib event rate is comparable to the SN IIb
rate~\citep[e.g.,][]{Smith11, Eldridge13, Shivvers16}.  

Note also that the optical brightness of SN IIb progenitors gradually decreases
as the surface temperature increases for $\log T_\mathrm{eff}/\mathrm{K} \gtrsim 3.9$.
This is mainly because a larger bolometric correction is needed for a higher
$T_\mathrm{eff}$. In particular,  SN IIb-B progenitors with $\log
T_\mathrm{eff}/\mathrm{K} \gtrsim 4.5$ have $M_\mathrm{V} > -5$ and would be relatively
difficult to detect in pre-SN images, compared to SN IIb-Y and IIb-R
progenitors.

\subsection{Final hydrogen mass - surface temperature relation}\label{sect:mhenv}

\begin{figure}
\begin{center}
\includegraphics[width =1.0\columnwidth]{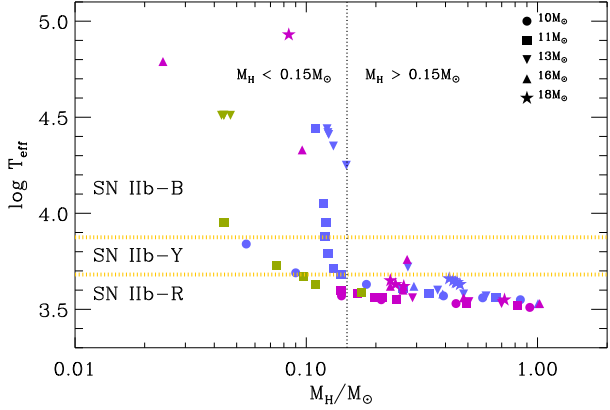}
\caption{
The final surface temperatures as a function of the hydrogen envelope mass ($M_\mathrm{H,env}$)
of our SN IIb progenitor models.  
The initial primary mass is marked by different symbols: circle, square, inverted triangle, triangle, and star
denote 10, 11, 13, 16 and 18~$\mathrm{M_\odot}$, respectively. 
Purple and blue indicate
solar and LMC metallicities, respectively. Green gives the result with 
a reduced mass loss rate (i.e., $\eta = 0.48$) at solar metallicity.  
The orange dotted lines mark the boundaries for yellow supergiants: 
$3.681 \le \log T_\mathrm{eff}/\mathrm{K} \le 3.875$.  
The back dotted vertical line roughly defines the boundary between extended progenitors (SN IIb-R), 
and relatively compact progenitors (SN IIb-Y and SN IIb-B), in terms of the hydrogen envelope mass. 
}\label{fig:mhteff}
\end{center}
\end{figure}

One of the dominant factors that determine the final size and surface
temperature of a SN IIb progenitor is the hydrogen envelope mass
($M_\mathrm{H,env}$) at the pre-SN stage. We find that SN IIb-R solutions are
dominant if $M_\mathrm{H,env} > 0.15~\mathrm{M_\odot}$, while SN IIb-Y and SN IIb-B
solutions are more common with a smaller hydrogen envelope mass
(Fig.~\ref{fig:mhteff}). This is in good agreement with the previous estimates
for the SNe IIb in Table~\ref{tab4}: $M_\mathrm{H,env} = 0.2 - 0.4~\mathrm{M_\odot}$
for SN 1993J and SN 2013df (RSG progenitor), $M_\mathrm{H,env} \approx
0.1~\mathrm{M_\odot}$ for SN 2011dh (YSG progenitor) and $M_\mathrm{H,env} \approx
0.06~\mathrm{M_\odot}$ for SN 2008ax (BSG progenitor). 

This result is also generally consistent with the finding of \citet{Meynet15},
who discussed single star SN progenitors at solar metallicity  with enhanced RSG mass
loss rates.  A few notable differences are also found as the following: 
\begin{enumerate}
\item In Meynet et al., SN progenitors with $\log T_\mathrm{eff,f}/\mathrm{K} > 3.6$ were found only with $M_\mathrm{H,env} < 0.4~\mathrm{M_\odot}$, 
but in our grid, this upper limit extends to $M_\mathrm{H,env} = 0.6~\mathrm{M_\odot}$. 
\item Meynet et al. find SN progenitors with  $\log T_\mathrm{eff,f}/\mathrm{K} < 3.6$ only for $M_\mathrm{H,env} > 0.9~\mathrm{M_\odot}$,
but in our study, solutions with  $\log T_\mathrm{eff,f}/\mathrm{K} < 3.6$ are  obtained even for smaller hydrogen masses  ($M_\mathrm{H,env} \gtrsim 0.2~\mathrm{M_\odot}$). 
\item In Meynet et al.,  no models with
$M_\mathrm{H,env} > 0.1~\mathrm{M_\odot}$ have a surface temperature higher than $\log T_\mathrm{eff,f}/\mathrm{K} = 4.3$.  
By contrast, many of our LMC metallicity models  have $\log T_\mathrm{eff,f}/\mathrm{K} >
4.3$ with $M_\mathrm{H,env}  = 0.1 -0.14~\mathrm{M_\odot}$. 
\item Our models include SN IIb-B progenitors with $\log T_\mathrm{eff,f}/\mathrm{K} \ge 4.8$ at solar metallicity,
which was not found in Meynet at al..  
\end{enumerate}
These differences can be explained by the fact that our models cover a wider parameter space than in Meynet et al., 
in terms of mass loss histories and metallicities. 

\subsection{Asymmetry of the progenitor envelope and winds}\label{sect:asymmetry}

\begin{figure}
\begin{center}
\includegraphics[width =1.0\columnwidth]{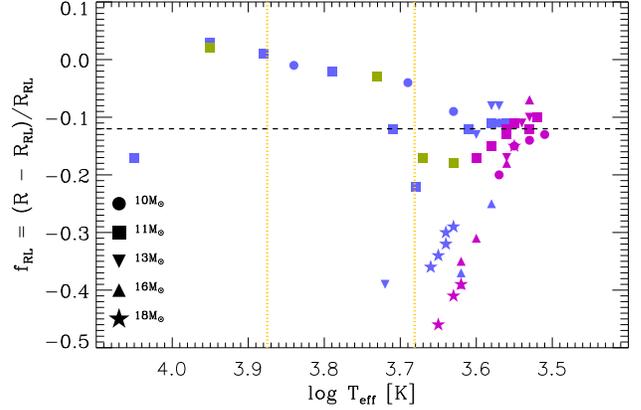}
\caption{The Roch-lobe filling factor for the last computed models of primary stars, 
defined by $f_\mathrm{RL} = (R - R_\mathrm{RL})/R_\mathrm{RL}$
where $R$ and $R_\mathrm{RL}$ denote the radius of the star and the Roche lobe radius, respectively. 
The orange dotted lines mark the boundaries for yellow supergiants: 
$3.681 \le \log T_\mathrm{eff}/\mathrm{K} \le 3.875$.  
}\label{fig:fRL}
\end{center}
\end{figure}

Compared to single stars, the outermost layers of SN progenitors in binary
systems can have an asymmetric structure at the pre-SN stage if they are
filling or almost filling the Roche lobe. Mass transfer and interaction between
stellar winds and transferred mass may also create complex, non-spherical
circumstellar  structures. These factors might be related to some polarization
features observed at early times in SNe IIb when the photosphere is
still located at the outermost ejecta layers (\citealt{Hoeflich95}; \citealt{Tran97};
\citealt{Maund07}; \citealt{Chornock11}; \citealt{Dessart11a}; \citealt{Mauerhan15};
\citealt{Stevance16}; cf. \citealt{Reilly16} for the SN Ib iPTF13bvn). 

We present the Roche lobe filling factor $f_\mathrm{RL}$ for our SN IIb
progenitor models in Fig.~\ref{fig:fRL}, which is defined as $f_\mathrm{RL} =
(R - R_\mathrm{RL})/R_\mathrm{RL}$, where $R_\mathrm{RL}$ is the Roche lobe
radius. Note that mass transfer becomes significant ($\dot{M}_\mathrm{tr}
\gtrsim 10^{-6}~\mathrm{M_\odot~yr^{-1}}$) if $f_\mathrm{RL} \gtrsim -0.12$,
according to the adopted mass transfer scheme in our calculations (see
Section~\ref{sect:method}).  Many of our SN IIb progenitor models end their lives
as a detached binary, but not a small fraction of the sequences have
$f_\mathrm{RL} > -0.12$ at the final stage, in particular for SN IIb-Y
progenitors with $M_\mathrm{1, init}  \le  11~\mathrm{M_\odot}$.  On the other
hand, all SN IIb-B progenitors with $\log T_\mathrm{eff}/\mathrm{K} \gtrsim 4.0$ are
detached from the Roche lobe because they are compact.  SN IIb-R progenitors
tend to be more detached with a higher mass  and no progenitors with
$M_\mathrm{1, init} = 18~\mathrm{M_\odot}$ fills the Roche lobe at the pre-SN stage.
This is because stronger wind mass loss from a more massive progenitor tends to make the
orbit wider than the corresponding case with a less massive progenitor.   

Whether or not the Roche lobe filling envelope at the pre-SN stage can explain
some features of polarization during an early SN epoch is an important subject
of future work.  Previous spectropolarimetry studies on core collapse SNe imply
that polarisation features are strongly connected to asymmetric
explosion ~\citep[e.g.,][]{Wang08}.  Although signatures of asymmetric
explosions are more prominent in late-time observations, its impact on shock
heating of the envelope and/or asymmetric distribution of nickel may also play
an important role in the early-time SN evolution~\citep{Mauerhan15}. 
Therefore, the distorted shape of the envelope would not be the only 
possible explanation for early-time polarisation.

The wind mass loss rates from SN IIb progenitors range from $\log
\dot{M}_\mathrm{w} [\mathrm{M_\odot~yr^{-1}}]  \simeq -6.5 $ to  $\log
\dot{M}_\mathrm{w} [\mathrm{M_\odot~yr^{-1}}]  \simeq -4.8$, and increase with
final mass and radius (see Tables~\ref{tab1},~\ref{tab2}
and~\ref{tab3}).  These mass loss rates are largely consistent with the values
inferred from observations of SN IIb interactions with circumstellar
medium~\citep[e.g.,][]{Maeda14, Maeda15}. 

The mass transfer rates from the Roche-lobe filling progenitors 
largely depend on $f_\mathrm{RL}$ at the final
stage, rather than on the final mass (see Tables~\ref{tab1}, \ref{tab2} and
\ref{tab3}).  In our assumption of non-conservative mass transfer, the
transferred material that is not accreted onto the secondary star is supposed to
be blown away from the binary system as a fast wind from the secondary star. On
the other hand,  the winds from Roche-lobe filling progenitors would be much
slower than the winds from secondary stars.  Even if the primary star is not
filling the Roche lobe, slow and fast winds are expected from a SN IIb-Y/IIb-R
progenitor and its dwarf companion, respectively.  The resultant wind-wind
collisions would make a complicated, clumpy circumstellar structure.  This
possibility and its observational signatures deserve future numerical
studies.

\section{Properties of SN I\lowercase{b} progenitors}\label{sect:snIb}

\begin{figure*}
\begin{center}
\includegraphics[width =1.0\columnwidth]{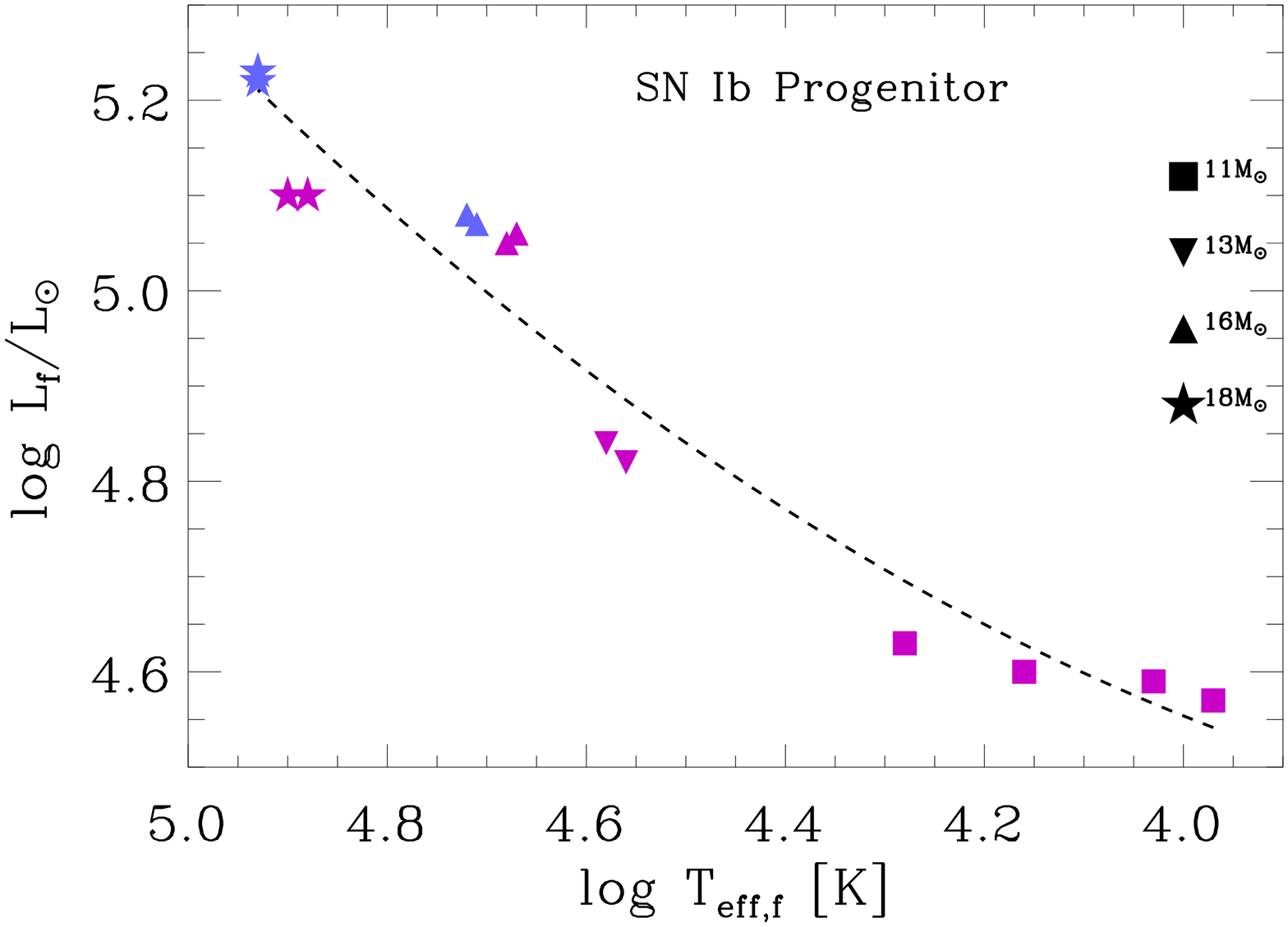}
\includegraphics[width =1.0\columnwidth]{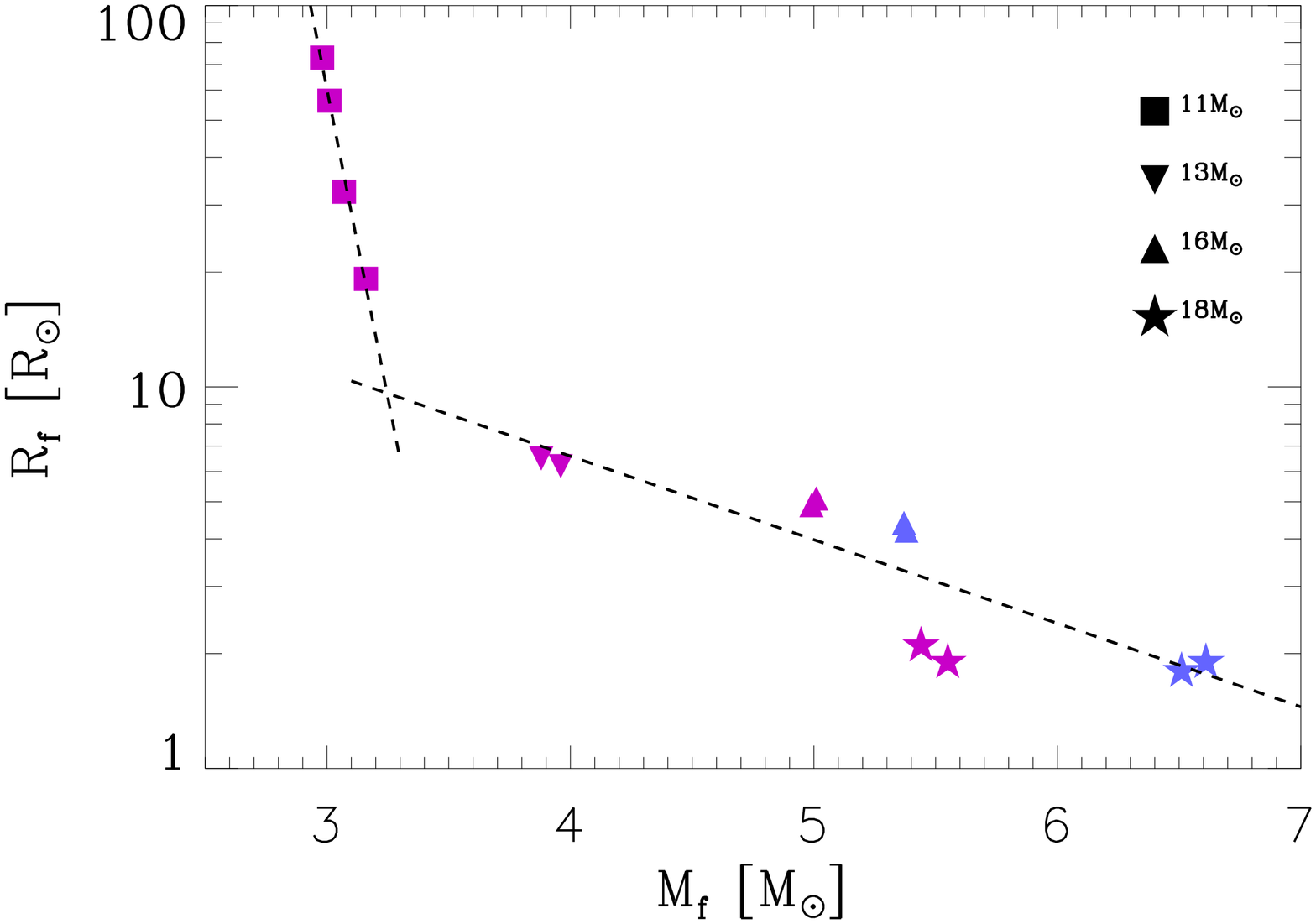}
\includegraphics[width =1.0\columnwidth]{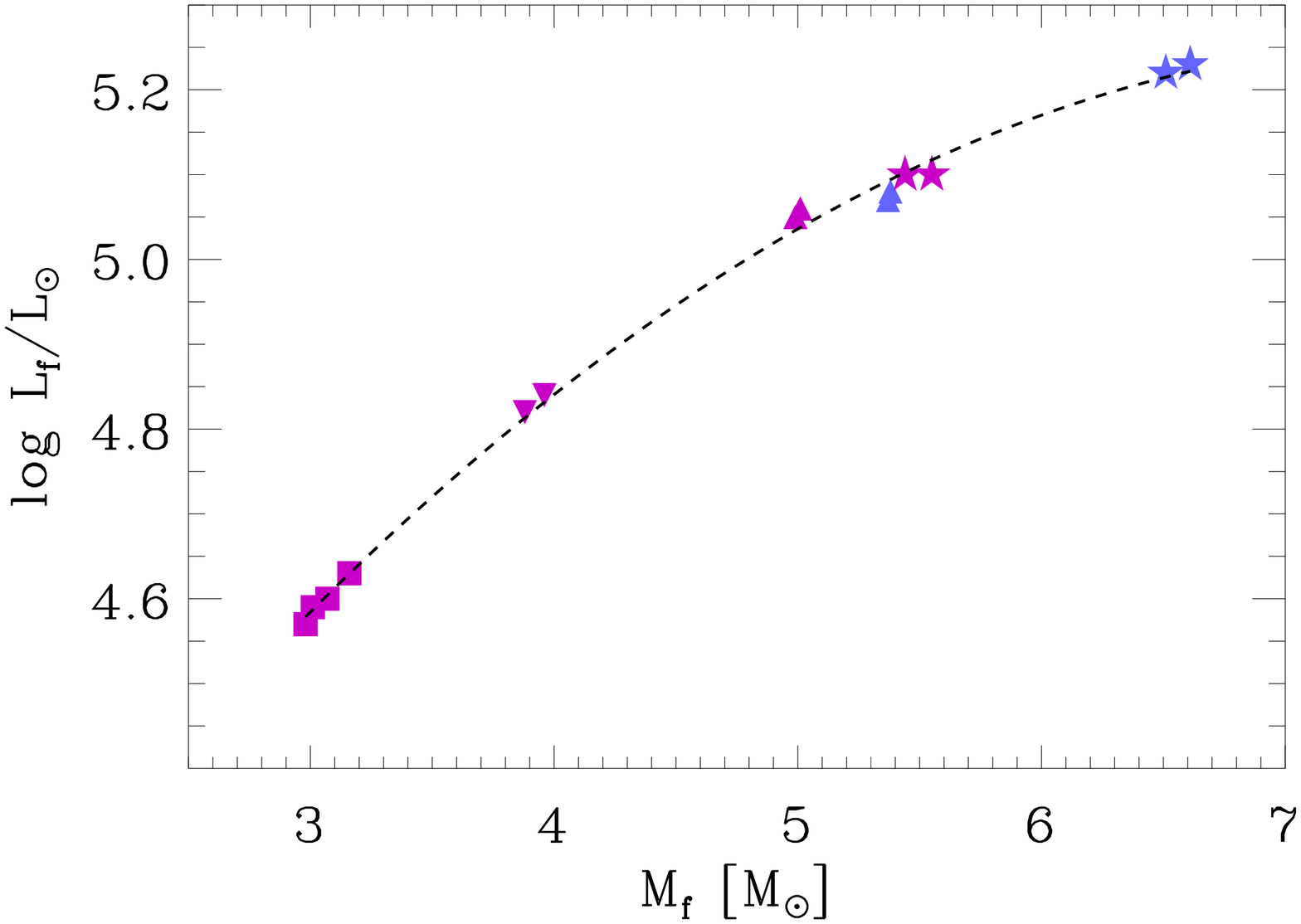}
\includegraphics[width =1.0\columnwidth]{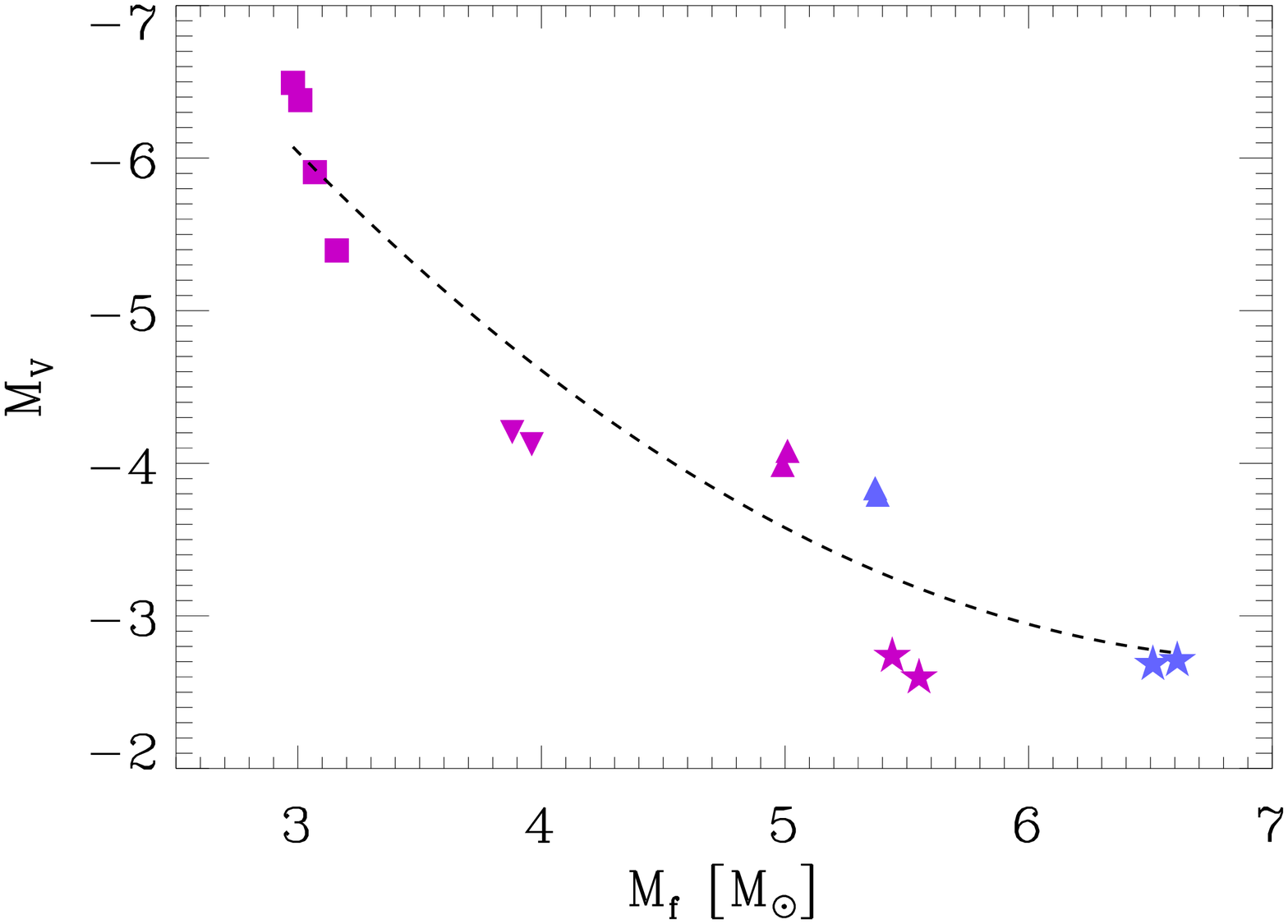}
\caption{\emph{Top left:} The final positions of our SN Ib progenitor models on the HR diagram. The different symbols
indicate the corresponding initial masses, as indicated by the associated numbers. LMC and solar metallicity models
are marked by blue and purple colors, respectively. \emph{Top right:} The final 
mass-radius relation of our SN Ib progenitor models. \emph{Bottom left:} The final mass-luminosity relation of 
our SN Ib progenitor models.  \emph{Bottom right:} The predicted optical magnitude in V-band of SN Ib progenitors
as a function of the final mass. 
}\label{fig:snIb}
\end{center}
\end{figure*}

Many of our considered systems produce SN Ib progenitors (mostly via Case EB
mass transfer).   All of our models contain a fairly large amount of helium in
the envelope ($m_\mathrm{He} > 1.5~\mathrm{M_\odot}$;
Tables~\ref{tab1},~\ref{tab2} and~\ref{tab3}), and the final masses of our
models are not large enough to hide this amount of helium in the
spectra~\citep[cf.][]{Hachinger12, Dessart12}.   Therefore no SN Ic is robustly
predicted within our grid.  The final mass ranges from 3.0~$\mathrm{M_\odot}$
to 6.6~$\mathrm{M_\odot}$, which can explain well the ejecta masses of most SNe
Ib ($M_\mathrm{ejecta} \simeq 1.0 - 5.0~\mathrm{M_\odot}$) inferred from
observations~\citep[e.g.,][]{Drout11, Taddia15, Lyman16} \footnote{Note,
however, that these ejecta masses have been inferred using a simplified model
based on \citet{Arnett82}, which is not very suitable for SNe
Ibc~\citep[see][]{Dessart16}.}. 

We present the final positions of our SN Ib progenitor models on the HR
diagram, in Fig.~\ref{fig:snIb}.  The final mass-radius and  mass-luminosity
relations are also shown in this figure.  Unlike SN IIb progenitors, there
exists a clear correlation between the final luminosity and the final
surface temperature.  More luminous (hence more massive) SN Ib progenitors have
higher surface temperatures.

This correlation between the final luminosity and surface temperature is
reflected by the final mass-radius relation shown in the top-right panel of the
figure.  The expansion of the helium envelope during the final evolutionary
stages becomes more significant for a more compact carbon-oxygen core (the
so-called mirror effect). This property of SN Ib/c progenitors was first
reported by \citet{Yoon10}, and our new models confirm their result, even
though we adopted a different mass loss rate prescription as discussed in
Section~\ref{sect:massloss} above.   This leads to the following
counter-intuitive result that was also discussed by \citet{Yoon12} in more
detail: the brightness in the optical increases as the progenitor mass
decreases, while the bolometric luminosity behaves in the opposite way.
This is because the bolometric correction increases with increasing 
$T_\mathrm{eff}$ (in this $T_\mathrm{eff}$ range). 
 
The expansion of the helium envelope is particularly strong when $M_\mathrm{f}
\lesssim 3.2~\mathrm{M_\odot}$.  The predicted optical brightness can reach
$M_\mathrm{V} = -5.4 \cdots -6.5$ for $M_\mathrm{f} \approx
3.0~\mathrm{M_\odot}$ (Fig.~\ref{fig:snIb}).  This may well explain the optical
brightness of the progenitor of iPTF13bvn, which was identified
by~\citet{Cao13}, even without a contribution from its companion, in good
agreement with the conclusion of previous studies~\citep{Kim15, Folatelli16,
Eldridge16}.  However, we need fine-tuned initial masses ($M_\mathrm{init} \sim
11~\mathrm{M_\odot}$) to get this bright progenitor in the optical.  More
massive progenitors are relatively faint  ($M_\mathrm{V} > - 5.0$), and would
be more difficult to detect unless they have a brighter companion star.  The
initial masses of these SN Ib progenitor models are in the range of $11 -
18~\mathrm{M_\odot}$.  In our assumption of $\beta = 0.2$ and $q=0.9$, the
companion star mass  does not exceed  about 18~$\mathrm{M_\odot}$, which cannot
be much brighter than $M_\mathrm{V} = -5.0$ at the pre-SN stage~\citep{Kim15}.
Therefore, a sufficiently nearby target would be needed
to detect a SN Ib progenitor with $3.5~\mathrm{M_\odot} \lesssim
M_\mathrm{f} \lesssim 7.0~\mathrm{M_\odot}$, unless the mass transfer is nearly
conservative.

\section{Metallicity effect}\label{sect:metallicity}

As shown in Fig.~\ref{fig:result}, the parameter space for SN Ib progenitors
broadens as the metallicity increases. This is because strong winds at a
high metallicity during the post mass transfer phase can completely remove the
remaining hydrogen layer, for sufficiently short initial orbits.  This implies
that the SN Ib to SN IIb event rate ratio  would increase with increasing
metallicity. This prediction seems to be supported by recent analyses on SN
host galaxies~(see Figs. A1 and A2 of \citealt{Graur16}; cf.\citealt{Arcavi10}), but
observational studies still suffer from low-statistics on SNe from metal poor
environments. The recent analysis on SN IIP by \citet{Dessart14} indicates
that most SN events occur at around solar metallicity. 

A few remarks should be made here.  
First, given that we still do
not have a good constraint on the mass loss rate of stars with stripped envelopes of
the relevant mass range as discussed in Section~\ref{sect:massloss},  our prediction
on the metallicity effect is qualitatively sound but quantitatively uncertain. 
Second,  our considered parameter
space is limited by  certain assumptions chosen for the present study that
focuses on stable mass transfer systems (i.e., a fixed mass ratio of $q=0.9$
and a rather low mass accretion efficiency of $\beta = 0.2$). Although the
binary channel with stable mass transfer is the most important path toward SN
Ib and SN IIb~\citep{Podsiadlowski93}, the role of non-stable mass transfer
needs to be investigated for completeness. Contribution from single
stars might also be significant, especially at super-solar
metallicity~\citep[cf.][]{Meynet15, Yoon15}.

We emphasize that this metallicity effect on the SN Ib/SN IIb ratio is
largely limited to the case of relatively compact progenitors (SN IIb-B and SN
IIb-Y), as clearly implied by Fig.~\ref{fig:result}. The regime for SN IIb-R is
not significantly affected by metallicity.  Therefore, a proper investigation
on the metallicity effect should distinguish sub-classes of SN IIb. 

More specifically, the following predictions on SN binary progenitors can be made with our result of Fig.~\ref{fig:result}.  
\begin{itemize}
\item The SN Ib/SN IIb-B and SN Ib/SN IIb-Y ratios would increase more rapidly than the SN Ib/SN IIb-R ratio, with
increasing metallicity. 
\item  The SN IIb-B/SN IIb-R and SN IIb-Y/SN IIb-R ratios would decrease with increasing metallicity. 
\item In  sufficiently metal-poor environments ($Z \approx 0.35Z_\odot$ with our fiducial mass loss rate), SN Ib
progenitors would be systematically more massive than SN IIb-B/SN IIb-Y
progenitors.  They would be also more massive than SN Ib progenitors from metal-rich environments, on average.  
\end{itemize}

\section{Conclusions}\label{sect:conclusions}

We have presented a new grid of models for SN Ib/IIb progenitors in binary systems.
We considered two different metallicities (solar and LMC metallicities) and the
initial masses of 10 -- 18~$\mathrm{M_\odot}$, which correspond to the final
masses of about 3.0 -- 8.0~$\mathrm{M_\odot}$. The initial mass ratio and mass
accretion efficiency were fixed to 0.9 and 0.2, respectively.  

Our investigation leads to the following conclusions. 
\begin{enumerate}
\item In general, a tighter initial orbit leads to stronger stripping of the hydrogen envelope
from the primary star. As a result, SN IIb progenitors 
from Case EB mass transfer in a relatively tight orbit are expected to have
compact sizes like SN 2008ax and SN 2011dh progenitors (blue and
yellow supergiants; SN IIb-B and SN IIb-Y), while Case LB mass transfer in a wider orbit usually
leads to red supergiant progenitors as in the case of SN 1993J and SN 2013df (SN
IIb-R; Fig.~\ref{fig:result}).  
\item SN IIb progenitors from Case EB mass transfer can be
produced only if mass loss during the post-mass transfer phase is sufficiently
weak (Section~\ref{sect:massloss}; Fig.~\ref{fig:result}).  This means that SN IIb-B and SN IIb-Y
progenitors are favored by relatively low initial masses and/or low
metallicities. SN IIb-R progenitors are expected for all initial masses and
metallicities considered in our study (Fig.~\ref{fig:result}).  
\item The
hydrogen envelope mass at the final stage is predicted to be higher than about
$M_\mathrm{H,env} = 0.15~\mathrm{M_\odot}$ in SN IIb-R progenitors, while  SN IIb-Y
and SN IIb-B progenitors would have a lower $M_\mathrm{H,env}$ (Fig.~\ref{fig:mhteff}). This prediction is
in good agreement with the previous estimates of $M_\mathrm{H,env} \simeq
0.1~\mathrm{M_\odot}$ for SN 2011dh (SN IIb-Y), and $M_\mathrm{H,env} \simeq 0.2 -
0.4~\mathrm{M_\odot}$ for SN 1993J and SN 2013df (SN IIb-R).  
\item In our
considered mass range, SN IIb-R progenitors would have $\log L/\mathrm{L_\odot}
\simeq 4.50 - 5.30$.  The luminosity range of SN IIb-B/IIb-Y progenitors is
somewhat narrower ($\log L/\mathrm{L_\odot} \simeq 4.51 - 5.15$),  and most of
them (i.e., SN IIb-B/II-Y from Case EB mass transfer) would have $\log
L/\mathrm{L_\odot} < 5.0$.  
\item The optical brightness of SN IIb progenitors
($M_\mathrm{V} < -5.0$ for most cases; Fig.~\ref{fig:hr}) would be systematically much higher than
SN Ib progenitors ($M_\mathrm{V} > -5.0$ for most cases; Fig.~\ref{fig:snIb}).  
For a given initial mass,  SN Ib progenitors have
higher surface temperatures and lower final
masses than SN IIb progenitors.
\item SN
Ib progenitors in binary systems are systematically hotter for higher masses.
As a result,  SN Ib progenitors become brighter in the optical for lower
masses, even though the bolometric luminosity increases with increasing mass
(Fig.~\ref{fig:snIb}).  The optically bright progenitor of SN Ib iPTF13bvn
($M_\mathrm{V} \sim -6$) may be well explained by a progenitor with a final
mass of about 3.0~$\mathrm{M_\odot}$ (Section~\ref{sect:snIb}).  
\item The impact of  wind mass loss after the mass transfer phase
becomes more significant at higher metallicity. As a result,  
the  SN Ib to SN IIb event rate ratio would increase with metallicity, while
the SN IIb-B/SN IIb-R and SN IIb-Y/SN IIb-R ratios would decrease (Section~\ref{sect:metallicity}).
By contrast, the parameter space for SN IIb-R progenitors is not significantly affected by metallicity 
for our considered mass range. 
\end{enumerate}

\clearpage

\begin{deluxetable*}{l|ccrrrrrrrrrrrcc}
\tabletypesize{\footnotesize}
\tablewidth{0pt}
\tablecaption{Physical properties of the final models of the primary star at $Z=0.007$.}\label{tab1}
\tablehead{
\colhead{Name} & \colhead{$M_\mathrm{f}$} & \colhead{$L_\mathrm{f}$}  &  \colhead{$R_\mathrm{f}$} &   \colhead{$T_\mathrm{eff, f}$}  
& \colhead{He} &  \colhead{CO} &  \colhead{H$\mathrm{env}$}  & \colhead{$m_\mathrm{H}$} &  
\colhead{$m_\mathrm{He}$} & \colhead{$Y_\mathrm{s, f}$} &  \colhead{$\dot{M}_\mathrm{tr}$} & 
\colhead{$\dot{M}_\mathrm{w}$} & \colhead{$f_\mathrm{RL}$}  & \colhead{Case} &  \colhead{SN}
}   
\startdata
Lm10p20  & 2.99 & 4.58 & 66.6 & 4.00 &   2.944  &  1.509 &  0.046 & 0.005  & 1.440 & 0.78 & -12.67  & -6.75 &    -0.15  & EBB & IIb-B \\
Lm10p50  & 3.01 & 4.61 & 143.1 & 3.84 &   2.955  &  1.511 &  0.055 & 0.007  & 1.448 & 0.76 & -5.51  & -6.44 &    -0.01  & EBB & IIb-Y \\
Lm10p150 & 3.07 & 4.61 & 278.8 & 3.69 &   2.979  &  1.527 &  0.090 & 0.015  & 1.486 & 0.73 & -5.33  & -6.19 &   -0.04  & EBB & IIb-Y\\
Lm10p300 & 3.33 & 4.65 & 387.4 & 3.63 &   3.147  &  1.602 &  0.182 & 0.048  & 1.622 & 0.57 & -6.44  & -6.02 &   -0.09  & EBB & IIb-Y\\
Lm10p700 & 3.70 & 4.69 & 529.6 & 3.57 &   3.311  &  1.676 &  0.392 & 0.205  & 1.745 & 0.38 & -6.04  & -5.85 &   -0.11  & LBB & IIb-R\\
Lm10p800 & 3.90 & 4.69 & 560.9 & 3.56 &   3.320  &  1.681 &  0.579 & 0.336  & 1.802 & 0.34 & -5.86  & -5.83 &   -0.11  & LBB & IIb-R\\
Lm10p900 & 4.18 & 4.69 & 583.2 & 3.55 &   3.335  &  1.686 &  0.843 & 0.526  & 1.889 & 0.31 & -5.71  & -5.81 &   -0.11  & LBB & IIb-R\\
Lm10p1000& 4.81 & 4.70 & 603.2 & 3.55 &   3.378  &  1.701 &  1.432 & 0.951  & 2.070 & 0.29 & -5.45  & -5.79 &  -0.09  & LBB & IIb-R\\
\tableline
Lm11p10   & 3.53 & 4.72 & 10.0 & 4.44 &   3.417  &  1.748 &  0.110 & 0.010  & 1.683 & 0.85 & -99.00  & -6.11 &    -0.79  & EB & IIb-B\\
Lm11p20  & 3.56 & 4.73 & 61.1 & 4.05 &   3.444  &  1.762 &  0.119 & 0.017  & 1.697 & 0.67 & -11.13  & -6.28 &    -0.17  & EB & IIb-B \\
Lm11p30  & 3.57 & 4.73 & 99.3 & 3.95 &   3.452  &  1.766 &  0.122 & 0.019  & 1.700 & 0.67 & -5.36  & -6.40 &      0.03  & EBB & IIb-B \\
Lm11p50  & 3.59 & 4.73 & 135.9 & 3.88 &   3.466  &  1.772 &  0.120 & 0.021  & 1.702 & 0.67 & -5.39  & -6.29 &     0.01  & EBB & IIb-B\\
Lm11p100  & 3.60 & 4.73 & 206.3 & 3.79 &   3.475  &  1.778 &  0.124 & 0.024  & 1.706 & 0.67 & -5.87  & -6.13 &   -0.02  & EBB & IIb-Y\\
Lm11p200  & 3.62 & 4.74 & 290.9 & 3.71 &   3.489  &  1.784 &  0.131 & 0.028  & 1.717 & 0.67 & -10.39  & -6.01 &  -0.12  & EB  & IIb-Y\\
Lm11p300  & 3.67 & 4.74 & 334.7 & 3.68 &   3.526  &  1.802 &  0.141 & 0.035  & 1.738 & 0.67 & -15.88  & -5.95 &  -0.22  & EB  & IIb-Y\\
Lm11p500  & 3.93 & 4.77 & 487.8 & 3.61 &   3.664  &  1.880 &  0.263 & 0.089  & 1.848 & 0.49 & -7.13  & -5.77 &   -0.12  & LBB & IIb-R \\
Lm11p700  & 4.08 & 4.79 & 561.6 & 3.58 &   3.741  &  1.919 &  0.341 & 0.160  & 1.888 & 0.40 & -5.92  & -5.70 &   -0.11  & LBB & IIb-R\\
Lm11p1000 & 4.45 & 4.79 & 630.6 & 3.56 &   3.787  &  1.988 &  0.660 & 0.388  & 1.991 & 0.33 & -5.66  & -5.65 &  -0.11  & LBB & IIb-R\\
Lm11p1200 & 5.03 & 4.80 & 668.6 & 3.55 &   3.811  &  1.957 &  1.220 & 0.793  & 2.146 & 0.29 & -5.36  & -5.62 &  -0.09  & LBB & IIL/IIP\\
\tableline
Lm13p20 &  4.42 & 4.90 & 12.6 & 4.44 &   4.295  &  2.273 &  0.123 & 0.017  & 1.958 & 0.72 & -99.00  & -6.00 &    -0.82  & EB & IIb-B\\
Lm13p50 &  4.42 & 4.90 & 13.8 & 4.42 &   4.299  &  2.276 &  0.124 & 0.017  & 1.972 & 0.72 & -99.00  & -6.01 &    -0.89  & EB & IIb-B\\
Lm13p100&  4.43 & 4.90 & 14.6 & 4.41 &   4.303  &  2.278 &  0.125 & 0.018  & 1.972 & 0.71 & -99.00  & -6.01 &   -0.93  & EB & IIb-B\\
Lm13p300&  4.46 & 4.91 & 19.1 & 4.35 &   4.329  &  2.295 &  0.131 & 0.021  & 1.981 & 0.71 & -99.00  & -6.00 &   -0.96  & EB & IIb-B\\
Lm13p400&  4.56 & 4.92 & 31.0 & 4.25 &   4.409  &  2.348 &  0.149 & 0.029  & 2.010 & 0.71 & -99.00  & -5.99 &   -0.94  & EB & IIb-B\\
Lm13p500&  4.80 & 4.94 & 350.6 & 3.72 &   4.521  &  2.415 &  0.275 & 0.066  & 2.128 & 0.70 & -37.90  & -5.66 &  -0.39  & EB & IIb-Y\\
Lm13p600&   contact \\
Lm13p900&   contact \\
Lm13p1000& 5.06 & 4.97 & 637.4 & 3.60 &   4.691  &  2.524 &  0.370 & 0.141  & 2.191 & 0.47 & -7.33  & -5.41 &  -0.13  & LBB & IIb-R \\
Lm13p1100& 5.20 & 4.97 & 690.3 & 3.58 &   4.723  &  2.543 &  0.479 & 0.223  & 2.231 & 0.42 & -5.21  & -5.38 &  -0.08  & LBB & IIb-R\\
Lm13p1300& 5.35 & 4.97 & 726.7 & 3.57 &   4.750  &  2.563 &  0.599 & 0.311  & 2.279 & 0.38 & -5.02  & -5.35 &  -0.08  & LBB & IIb-R\\
Lm13p1500& 5.87 & 4.98 & 794.0 & 3.56 &   4.774  &  2.578 &  1.097 & 0.667  & 2.417 & 0.32 & -4.72  & -5.31 &  -0.08  & LBB & IIb-R\\
\tableline
Lm16p50 &   5.37 & 5.07 & 4.4 & 4.71 &   5.368  &  3.143 &  0.000 & 0.000  & 2.041 & 0.99 & -99.00  & -5.55 &     -0.97  & EB & Ib\\
Lm16p100&   5.38 & 5.08 & 4.2 & 4.72 &   5.383  &  3.173 &  0.000 & 0.000  & 2.023 & 0.99 & -99.00  & -5.54 &    -0.98  & EB & Ib\\
Lm16p200&    contact\\
Lm16p500&     contact\\
Lm16p1800&    contact\\
Lm16p1900&    contact\\
Lm16p2000&  6.35 & 5.15 & 713.3 & 3.62 &   6.056  &  3.774 &  0.292 & 0.074  & 2.277 & 0.57 & -24.51  & -5.12 &    -0.37 & LB  & IIb-R \\
Lm16p2400&  6.48 & 5.15 & 865.3 & 3.58 &   6.086  &  3.657 &  0.389 & 0.129  & 2.491 & 0.52 & -13.10  & -5.05 &    -0.25 & LB  & IIb-R\\
Lm16p2600&  7.12 & 5.16 & 1098.3 & 3.53 &   6.116  &  3.679 &  1.005 & 0.553  & 2.678 & 0.37 & -4.51  & -4.95 &    -0.07 & LBB & IIL/IIP \\
\tableline
Lm18p50 &   6.51 & 5.22 & 1.8 & 4.93 &   6.508  &  4.221 &  0.000 & 0.000  & 2.116 & 0.99 & -99.00  & -5.36 &    -0.99   & EB & Ib \\
Lm18p300 &  6.61 & 5.23 & 1.9 & 4.93 &   6.614  &  4.276 &  0.000 & 0.000  & 2.172 & 0.99 & -99.00  & -5.35 &    -1.00  & EB & Ib \\
Lm18p400 &   contact \\
Lm18p1700&    contact \\
Lm18p1900&     contact \\
Lm18p2000&  7.73 & 5.29 & 710.8 & 3.66 &   7.311  &  4.707 &  0.415 & 0.111  & 2.707 & 0.59 & -26.18  & -4.93 &    -0.36  & LB & IIb-R  \\
Lm18p2100&  7.76 & 5.29 & 744.4 & 3.65 &   7.328  &  4.722 &  0.432 & 0.120  & 2.718 & 0.59 & -23.33  & -4.91 &    -0.34  & LB & IIb-R  \\
Lm18p2200&  7.79 & 5.29 & 769.8 & 3.64 &   7.348  &  4.738 &  0.443 & 0.127  & 2.731 & 0.59 & -21.32  & -4.90 &    -0.32  & LB & IIb-R  \\
Lm18p2300&  7.81 & 5.30 & 787.0 & 3.64 &   7.363  &  4.749 &  0.449 & 0.131  & 2.730 & 0.59 & -20.02  & -4.89 &    -0.30  & LB &  IIb-R  \\
Lm18p2400&  7.84 & 5.30 & 806.4 & 3.63 &   7.380  &  4.761 &  0.462 & 0.138  & 2.747 & 0.59 & -18.70  & -4.87 &    -0.29  & LB &  IIb-R \\
Lm18p2500&  8.46 & 5.30 & 1039.8 & 3.58 &   7.411  &  4.784 &  1.047 & 0.528  & 2.935 & 0.39 & -4.84  & -4.78 &    -0.07  & LBB & IIL/IIP \\
\enddata
\tablecomments{$M_\mathrm{f}$: total mass in units of $\mathrm{M}_\odot$, $L_\mathrm{f}$: luminosity in units of $\mathrm{L_\odot}$ in logarithmic scale, $R_\mathrm{f}$: radius in units of $\mathrm{R_\odot}$, $T_\mathrm{eff}$: surface temperature in units of K in logarithmic scale, He: He core mass, CO : carbon-oxygen core mass, H$_\mathrm{env}$ : mass of the hydrogen envelope, $m_\mathrm{H}$: total mass
of hydrogen (i.e., $m_\mathrm{H} = \int X_\mathrm{H} dM_r$), $m_\mathrm{He}$ : total mass of helium, $Y_\mathrm{s,f}$: mass fraction of helium
at the surface, $\dot{M}_\mathrm{tr}$: mass transfer rate in units of $\mathrm{M_\odot~yr^{-1}}$ in logarithmic scale, 
$\dot{M}_w$: wind mass loss rate in units of $\mathrm{M_\odot~yr^{-1}}$ in logarithmic scale, $f_\mathrm{RL}$: Roche-lobe filling factor.} 
\end{deluxetable*}

\begin{deluxetable*}{l|ccrrrrrrrrrrrcc}
\tabletypesize{\footnotesize}
\tablewidth{0pt}
\tablecaption{Physical properties of the final models of the primary star at $Z=0.02$.}\label{tab2}
\tablehead{
\colhead{Name} & \colhead{$M_\mathrm{f}$} & \colhead{$L_\mathrm{f}$}  &  \colhead{$R_\mathrm{f}$} &   \colhead{$T_\mathrm{eff, f}$}  
& \colhead{He} &  \colhead{CO} &  \colhead{H$\mathrm{env}$}  & \colhead{$m_\mathrm{H}$} &  
\colhead{$m_\mathrm{He}$} & \colhead{$Y_\mathrm{s, f}$} &  \colhead{$\dot{M}_\mathrm{tr}$} & 
\colhead{$\dot{M}_\mathrm{w}$} & \colhead{$f_\mathrm{RL}$}  & \colhead{Case} &  \colhead{SN}
}   
\startdata
Sm10p50  & 2.04 & 4.66 & 258.5 & 3.72 &   2.041  &  1.360 &  0.000 & 0.000  & 0.660 & 0.97 & -1.52  & -6.16 &      0.12  & EBB &  --  \\
Sm10p200 & 2.73 & 4.43 & 139.0 & 3.80 &   2.726  &  1.391 &  0.000 & 0.000  & 1.294 & 0.98 & -99.00  & -6.69 &   -0.64  & EB  &  --   \\
Sm10p400 & 3.12 & 4.52 & 429.0 & 3.57 &   2.976  &  1.501 &  0.142 & 0.047  & 1.507 & 0.54 & -11.29  & -6.16 &   -0.20  & LB  & IIb-R \\
Sm10p600 & 3.29 & 4.59 & 532.2 & 3.55 &   3.084  &  1.541 &  0.211 & 0.110  & 1.573 & 0.40 & -7.51  & -5.98 &    -0.15  & LB  & IIb-R \\
Sm10p800 & 3.55 & 4.61 & 598.5 & 3.53 &   3.105  &  1.554 &  0.444 & 0.262  & 1.656 & 0.34 & -6.54  & -5.91 &    -0.14  & LB  & IIb-R \\
Sm10p1000& 4.04 & 4.61 & 633.9 & 3.51 &   3.116  &  1.560 &  0.926 & 0.588  & 1.808 & 0.31 & -6.33  & -5.89 &   -0.13   \\
\tableline
Sm11p20 &  2.98 & 4.57 & 73.0 & 3.97 &   2.978  &  1.528 &  0.000 & 0.000  & 1.391 & 0.98 & -35.55  & -6.73 &    -0.15  & EB & Ib  \\
Sm11p50 &  3.01 & 4.59 & 56.3 & 4.03 &   3.005  &  1.539 &  0.000 & 0.000  & 1.404 & 0.98 & -99.00  & -6.04 &    -0.64  & EB & Ib\\
Sm11p200&  3.07 & 4.60 & 32.5 & 4.16 &   3.069  &  1.563 &  0.000 & 0.000  & 1.438 & 0.98 & -99.00  & -5.94 &    -0.92  & EB & Ib\\
Sm11p300&  3.16 & 4.63 & 19.2 & 4.28 &   3.158  &  1.598 &  0.000 & 0.000  & 1.487 & 0.98 & -99.00  & -5.90 &    -0.96  & EB & Ib\\
Sm11p400&  3.47 & 4.68 & 458.4 & 3.60 &   3.330  &  1.679 &  0.141 & 0.033  & 1.664 & 0.67 & -11.29  & -5.92 &   -0.17  & LB & IIb-R \\
Sm11p500&  3.55 & 4.69 & 511.9 & 3.58 &   3.387  &  1.709 &  0.166 & 0.052  & 1.689 & 0.53 & -8.58  & -5.86 &    -0.15  & LB & IIb-R\\
Sm11p600&  3.66 & 4.70 & 565.4 & 3.56 &   3.459  &  1.743 &  0.198 & 0.082  & 1.719 & 0.46 & -6.76  & -5.81 &    -0.12  & LB & IIb-R\\
Sm11p700&  3.71 & 4.71 & 586.9 & 3.56 &   3.497  &  1.765 &  0.212 & 0.096  & 1.732 & 0.41 & -6.57  & -5.78 &    -0.13  & LB & IIb-R\\
Sm11p800&  3.79 & 4.75 & 642.2 & 3.55 &   3.544  &  1.888 &  0.246 & 0.127  & 1.636 & 0.37 & -5.79  & -5.69 &    -0.11  & LBB & IIb-R \\
Sm11p1000& 4.04 & 4.72 & 675.4 & 3.53 &   3.547  &  1.790 &  0.495 & 0.285  & 1.840 & 0.34 & -6.08  & -5.72 &   -0.12  & LBB & IIb-R\\
Sm11p1200& 4.39 & 4.76 & 743.2 & 3.52 &   3.575  &  1.903 &  0.818 & 0.507  & 1.831 & 0.31 & -5.30  & -5.63 &   -0.10  & LBB & IIb-R\\
Sm11p1400& 5.62 & 4.73 & 732.9 & 3.51 &   3.606  &  1.816 &  2.018 & 1.316  & 2.323 & 0.30 & -5.75  & -5.67 &   -0.09  & LBB & IIL/IIP\\
\tableline
Sm13p50  &  3.88 & 4.82 & 6.5 & 4.56 &   3.878  &  2.095 &  0.000 & 0.000  & 1.627 & 0.98 & -99.00  & -5.65 &      -0.95 & EB & Ib  \\ 
Sm13p500 &  3.96 & 4.84 & 6.2 & 4.58 &   3.957  &  2.144 &  0.000 & 0.000  & 1.645 & 0.98 & -99.00  & -5.63 &     -0.99  & EB & Ib \\
Sm13p600 &     contact \\
Sm13p700 &       contact  \\
Sm13p1100&       contact  \\
Sm13p1150&       contact  \\
Sm13p1200&  4.86 & 4.93 & 743.2 & 3.56 &   4.567  &  2.419 &  0.288 & 0.133  & 2.078 & 0.38 & -8.04  & -5.40 &   -0.17  & LB & IIb-R \\
Sm13p1500 & 5.12 & 4.94 & 828.3 & 3.54 &   4.618  &  2.461 &  0.505 & 0.278  & 2.152 & 0.34 & -5.67  & -5.35 &   -0.11  & LB & IIb-R\\
Sm13p1700 & 5.33 & 4.94 & 866.6 & 3.53 &   4.629  &  2.461 &  0.699 & 0.407  & 2.225 & 0.33 & -5.28  & -5.33 &   -0.10  & LB & IIb-R \\
Sm13p2000 & 6.08 & 4.95 & 933.4 & 3.51 &   4.666  &  2.487 &  1.411 & 0.890  & 2.447 & 0.30 & -4.85  & -5.29 &   -0.09  & LB & IIL/IIP\\
\tableline
Sm16p50 &  4.99 & 5.05 & 4.9 & 4.68 &   4.994  &  3.182 &  0.000 & 0.000  & 1.659 & 0.98 & -99.00  & -5.35 &      -0.96 & EB & Ib \\
Sm16p300&  5.01 & 5.06 & 5.1 & 4.67 &   5.009  &  3.208 &  0.000 & 0.000  & 1.649 & 0.98 & -99.00  & -5.34 &     -0.99 & EB & Ib \\
Sm16p400&         contact \\
Sm16p1400&        contact \\
Sm16p1600&        contact \\
Sm16p1700& 6.08 & 5.14 & 3.2 & 4.79 &   6.056  &  3.644 &  0.024 & 0.001  & 2.250 & 0.94 & -99.00  & -5.27 &    -1.00  & LB & IIb-B  \\
Sm16p1800& 6.19 & 5.16 & 27.5 & 4.33 &   6.098  &  3.678 &  0.096 & 0.011  & 2.293 & 0.76 & -99.00  & -5.40 &   -0.98  & LB & IIb-B  \\
Sm16p1900& 6.43 & 5.12 & 364.4 & 3.76 &   6.157  &  3.700 &  0.273 & 0.073  & 2.496 & 0.62 & -99.00  & -5.39 &   -0.67 & LB & IIb-Y  \\
Sm16p2000& 6.42 & 5.18 & 748.1 & 3.62 &   6.189  &  3.725 &  0.231 & 0.059  & 2.418 & 0.62 & -24.11  & -5.07 &   -0.35 & LB & IIb-R  \\
Sm16p2200& 6.48 & 5.15 & 717.3 & 3.62 &   6.214  &  3.746 &  0.262 & 0.074  & 2.470 & 0.62 & -28.71  & -5.12 &   -0.39 & LB & IIb-R  \\
Sm16p2400& 6.53 & 5.18 & 822.2 & 3.60 &   6.267  &  3.893 &  0.262 & 0.078  & 2.300 & 0.62 & -19.33  & -5.03 &   -0.31 & LB & IIb-R  \\
Sm16p2600& 6.76 & 5.18 & 974.9 & 3.56 &   6.285  &  3.805 &  0.479 & 0.204  & 2.531 & 0.46 & -8.91  & -4.97 &    -0.18 & LB & IIb-R  \\
Sm16p2700& 7.33 & 5.18 & 1121.5 & 3.53 &   6.307  &  3.815 &  1.021 & 0.575  & 2.693 & 0.36 & -4.42  & -4.91 &   -0.07 & LBB& IIL/IIP  \\
\tableline
Sm18p50 & 5.44 & 5.10 & 2.1 & 4.88 &   5.440  &  3.478 &  0.000 & 0.000  & 1.571 & 0.98 & -99.00  & -5.29 &    -0.98 & EB & Ib   \\
Sm18p500& 5.55 & 5.10 & 1.9 & 4.90 &   5.548  &  3.536 &  0.000 & 0.000  & 1.613 & 0.98 & -99.00  & -5.29 &   -1.00 & EB & Ib  \\ 
Sm18p600&        contact \\
Sm18p1000&       contact \\
Sm18p1900&       contact \\
Sm18p2000&6.62 & 5.19 & 1.7 & 4.94 &   6.624  &  4.330 &  0.000 & 0.000  & 2.158 & 0.98 & -99.00  & -5.18 &    -1.00  & LB & Ib \\ 
Sm18p2200&7.04 & 5.16 & 1.7 & 4.93 &   6.953  &  4.411 &  0.084 & 0.007  & 2.526 & 0.81 & -99.00  & -5.36 &    -1.00  & LB & IIb-B \\
Sm18p2400&7.28 & 5.26 & 710.3 & 3.65 &   7.047  &  4.458 &  0.231 & 0.050  & 2.565 & 0.67 & -41.32  & -4.97 &    -0.46 & LB & IIb-R  \\
Sm18p2800&7.37 & 5.27 & 794.0 & 3.63 &   7.126  &  4.623 &  0.243 & 0.058  & 2.475 & 0.67 & -32.79  & -4.92 &    -0.41 & LB & IIb-R \\
Sm18p3000&7.43 & 5.27 & 818.4 & 3.62 &   7.170  &  4.594 &  0.264 & 0.066  & 2.561 & 0.67 & -30.71  & -4.91 &    -0.39 & LB & IIb-R  \\
Sm18p3100&7.92 & 5.27 & 1140.1 & 3.55 &   7.203  &  4.585 &  0.720 & 0.349  & 2.767 & 0.37 & -7.41  & -4.79 &    -0.15 & LB & IIb-R  \\
\enddata
\end{deluxetable*}

\begin{deluxetable*}{l|ccrrrrrrrrrrrcc}
\tabletypesize{\footnotesize}
\tablewidth{0pt}
\tablecaption{Physical properties of the final models of the primary star at $Z = 0.02$ with a reduced mass loss rate ($\eta = 0.48$)}\label{tab3}
\tablehead{
\colhead{Name} & \colhead{$M_\mathrm{f}$} & \colhead{$L_\mathrm{f}$}  &  \colhead{$R_\mathrm{f}$} &   \colhead{$T_\mathrm{eff, f}$}  
& \colhead{He} &  \colhead{CO} &  \colhead{H$\mathrm{env}$}  & \colhead{$m_\mathrm{H}$} &  
\colhead{$m_\mathrm{He}$} & \colhead{$Y_\mathrm{s, f}$} &  \colhead{$\dot{M}_\mathrm{tr}$} & 
\colhead{$\dot{M}_\mathrm{w}$} & \colhead{$f_\mathrm{RL}$}  & \colhead{Case} &  \colhead{SN}
}   
\startdata
Tm11p20&   3.21 & 4.63 & 87.4 & 3.95 &   3.161  &  1.591 &  0.044 & 0.004  & 1.534 & 0.81 & -5.27  & -6.91 &     0.02   & EBB & IIb-B    \\
Tm11p100&  3.27 & 4.64 & 237.4 & 3.73 &   3.195  &  1.611 &  0.074 & 0.009  & 1.567 & 0.77 & -5.62  & -6.53 &    -0.03 & EBB & IIb-Y  \\
Tm11p200&  3.33 & 4.65 & 314.9 & 3.67 &   3.231  &  1.629 &  0.097 & 0.015  & 1.603 & 0.76 & -14.33  & -6.42 &    -0.17 & EB & IIb-R \\
Tm11p300&  3.40 & 4.67 & 395.7 & 3.63 &   3.290  &  1.661 &  0.110 & 0.021  & 1.628 & 0.71 & -12.72  & -6.31 &    -0.18 & LB & IIb-R \\
Tm11p400&  3.58 & 4.70 & 503.0 & 3.59 &   3.408  &  1.720 &  0.173 & 0.049  & 1.707 & 0.59 & -6.76  & -6.18 &    -0.10  & LBB & IIb-R \\
Tm11p500&   contact \\
\tableline
Tm13p20 &  4.26 & 4.87 & 8.6 & 4.51 &   4.218  &  2.183 &  0.044 & 0.002  & 1.887 & 0.87 & -99.00  & -5.99 &    -0.89  & EB & IIb-B  \\
Tm13p50 &  4.27 & 4.87 & 8.6 & 4.51 &   4.230  &  2.191 &  0.043 & 0.002  & 1.888 & 0.87 & -99.00  & -5.99 &    -0.94  & EB & IIb-B \\
Tm13p100&  4.28 & 4.87 & 8.7 & 4.51 &   4.234  &  2.193 &  0.044 & 0.002  & 1.895 & 0.87 & -99.00  & -5.99 &    -0.96 & EB & IIb-B  \\
Tm13p300&  4.32 & 4.88 & 8.8 & 4.51 &   4.270  &  2.218 &  0.047 & 0.003  & 1.908 & 0.86 & -99.00  & -5.99 &    -0.98 & EB & IIb-B   \\
Tm13p500&   contact \\
\tableline
Tm16p50 &  5.55 & 5.09 & 4.6 & 4.71 &   5.548  &  3.242 &  0.000 & 0.000  & 2.095 & 0.98 & -99.00  & -5.62 &    -0.97  & EB & Ib  \\
Tm16p100&  5.57 & 5.09 & 4.6 & 4.70 &   5.572  &  3.253 &  0.000 & 0.000  & 2.110 & 0.98 & -99.00  & -5.62 &    -0.98 & EB & Ib  \\
\enddata
\end{deluxetable*}

\begin{table}
\begin{center}
\caption{Properties of the identified SN IIb progenitors}\label{tab4}
\begin{tabular}{l c c c l}
\hline
          &   $\log L/\mathrm{L_\odot}$ & $\log T_\mathrm{eff}$ & $M_\mathrm{H,env}/\mathrm{M_\odot}$  & Ref.  \\
\hline 
SN 1993J  &       5.1$\pm$0.3           & 3.63$\pm$0.05         & 0.2 -- 0.4             & (1)  \\
SN 2008ax &      4.41 --  5.30          & 3.88 -- 4.30          & 0.06                   & (2)  \\
SN 2011dh &      4.90 -- 4.99           & 3.779$\pm$0.02        & 0.1                    & (3) \\
SN 2013df &      4.94$\pm$0.06          & 3.628$\pm$0.01        & 0.2 -- 0.4              & (4)  \\
SN 2016gkg &     4.65 -- 5.32           & 3.978$^{+0.215}_{-0.158}$  &    -             & (5)  \\
\hline
\end{tabular}
\end{center}
{(1)\citet{Aldering94, Maund04},  (2)\citet{Folatelli15},  (3)\citet{Maund11, VanDyk11, Bersten12},  (4)\citet{VanDyk14}, (5)\citet{Kilpatrick16, Tartaglia16} }
\end{table}

\section*{Acknowledgements}
This work was supported by the Korea Astronomy and Space Science Institute
under the R\&D program (Project No. 3348- 20160002) supervised by the Ministry
of Science, ICT and Future Planning.  AC acknowledges support from the Ministry
of Economy, Development, and Tourism's Millennium Science Initiative through
grant IC120009, awarded to The Millennium Institute of Astrophysics, MAS, and
from grant Basal CATA PFB 06/09.

\end{document}